%% file: main.tex
\documentclass[journal,compsoc]{IEEEtran}
\usepackage{amsmath,amssymb,amsfonts}
\usepackage{algorithm}
\usepackage{algorithmicx}
\usepackage{algpseudocode}
\usepackage{graphicx}
\usepackage{xcolor}
\usepackage{caption}
\usepackage{listings}
\usepackage{multirow}
\usepackage{multicol}
\usepackage{makecell}
\usepackage{xcolor, soul}
\usepackage{enumitem}
\usepackage{booktabs}
\usepackage{tabularx}
\usepackage{colortbl}
\usepackage{tcolorbox}
\tcbuselibrary{skins}
\usepackage{xspace}
\usepackage{pbox}
\usepackage[all]{nowidow}
\usepackage[numbers]{natbib}

\newcolumntype{Y}{>{\centering\arraybackslash}X}
\newcommand{\TOOL}{Paska\xspace}

\input{revision_1/commands}

\newcommand\revision[1]{\textcolor{black}{#1}}

\algnewcommand\algorithmicforeach{\textbf{for each}}
\algdef{S}[FOR]{ForEach}[1]{\algorithmicforeach\ #1\ \algorithmicdo}

\algnewcommand\algorithmicinput{\textbf{Input:}}
\algnewcommand\Input{\item[\algorithmicinput]}
\algnewcommand\algorithmicoutput{\textbf{Output:}}
\algnewcommand\Output{\item[\algorithmicoutput]}
\algrenewcommand\algorithmiccomment[1]{\State\textcolor{gray}{// #1}}

\algdef{SE}[REPEATN]{RepeatN}{EndRepeatN}[1]{\algorithmicrepeat\ #1 \textbf{times}}{\algorithmicend}

\begin{document}

\title{Automated Smell Detection and Recommendation in Natural Language Requirements}

\author{Alvaro Veizaga, Seung Yeob Shin,~\IEEEmembership{Member,~IEEE,} Lionel C. Briand,~\IEEEmembership{Fellow,~IEEE}
\IEEEcompsocitemizethanks{
\IEEEcompsocthanksitem Alvaro Veizaga is with the Interdisciplinary Centre for Security, Reliability, and Trust (SnT) of the University of Luxembourg, Luxembourg (e-mail: alvaro.veizaga@uni.lu).
\IEEEcompsocthanksitem Seung Yeob Shin is with the Interdisciplinary Centre for Security, Reliability, and Trust (SnT) of the University of Luxembourg, Luxembourg (e-mail: seungyeob.shin@uni.lu).
\IEEEcompsocthanksitem Lionel C. Briand holds shared appointments with the Interdisciplinary Centre for Security, Reliability, and Trust (SnT) of the University of Luxembourg, Luxembourg and the school of EECS, University of Ottawa, Ottawa, Canada (e-mail: lionel.briand@uni.lu).
}
}


\IEEEtitleabstractindextext{%
\begin{abstract}
Requirement specifications are typically written in natural language (NL) due to its usability across multiple domains and understandability by all stakeholders. However, unstructured NL is prone to quality problems (e.g., ambiguity) when writing requirements, which can result in project failures.
\revision{To address this issue, we present a tool, named \TOOL, that takes as input any NL requirements, automatically detects quality problems as smells in the requirements, and offers recommendations to improve their quality.}
Our approach relies on natural language processing (NLP) techniques and a state-of-the-art controlled natural language (CNL) for requirements (Rimay), to detect smells and suggest recommendations using patterns defined in Rimay to improve requirement quality. We evaluated \TOOL through an industrial case study in the financial domain involving 13 systems and 2725 annotated requirements. The results show that our tool is accurate in detecting smells (89\% precision and recall) and suggesting appropriate Rimay pattern recommendations (96\% precision and 94\% recall).
\end{abstract}

}

\input{CNL.tex}

\setcounter{table}{0}
\setcounter{algorithm}{0}
\setcounter{figure}{0}
\twocolumn

\maketitle
\setcounter{page}{1}

\input{introduction}

\input{background}
\input{context}
\input{input}

\input{approach.tex}

\input{evaluation}

\input{discussion}

\input{threats}

\input{related}

\input{conclusions}

\section*{Acknowledgment}
This project was supported by FNR of Luxembourg under the BRIDGES program (grant BRIDGES18/IS/13234469/IMoReF) and NSERC of Canada under the Discovery and CRC programs.


\bibliographystyle{IEEEtranN}  
\bibliography{ref}  

\end{document}

%% file: revision_1/commands.tex
\newcounter{commentnumber}
\newcounter{metacommentnumber}

%% file: CNL.tex
\definecolor{codegreen}{rgb}{0,0.6,0}
\definecolor{codegray}{rgb}{0.5,0.5,0.5}
\definecolor{codepurple}{rgb}{0.58,0,0.82}
\definecolor{backcolour}{rgb}{0.95,0.95,0.92}
\definecolor{bluegreen}{RGB}{51,153,126}	
\definecolor{intenttypecolor}{RGB}{58,162,187}
\definecolor{key-color}{rgb}{0.8, 0.47, 0.196}
\definecolor{xtext-keyword}{RGB}{127, 0, 85}
\definecolor{javadocblue}{rgb}{0.25,0.35,0.75} 

\newcommand{\CNL}{\lstinline[language=OwnExamples, basicstyle=\normalsize\ttfamily,
showstringspaces=false,
breakatwhitespace=false,
keepspaces=true,
breaklines=true,
prebreak=\mbox{},
postbreak=\mbox{} 
]}

\newcommand{\tinyCNL}{\lstinline[language=OwnExamples, basicstyle=\footnotesize\ttfamily,
breaklines=true, 
keepspaces=true,
showstringspaces=false,
breakatwhitespace=true
]}

\newcommand{\exCNL}{\lstinline[language=OwnExamples, basicstyle=\small\ttfamily,
breaklines=true,
keepspaces=true,
showstringspaces=false,
breakatwhitespace=true
]}

\lstdefinelanguage{OwnExamples}{
  keywords={all, each,zero,one,any,none, (, ),
     ,, ,_, ,_and, ,_or, -, ., :, Actors:, After, Every, every, Before, For, If, Quantifiers:, The, Types:, UI_Component_Types:, Units:, When, Where, While, _and, _or, a, about, accept, accepts, actions, actor, add, adds, after, allow, allows, an, and, and_, any_value, applies, apply, applying, are, as, available, based, based_on, before, by, calculate, calculates, checking, clicks, compliance, complies, comply, component, conditions, conform, conforms, contain, contains, convention, copies, copy, create, creates, deduct, deducts, default, defined, delete, deletes, described, detect, detects, disable, display, displayed, displays, do, does, double, download, downloads, during, enable, entity, entity_named, equal, equals, every, exclude, excludes, find, finds, finish, finishes, focuses, following, for, format, from, greater, has, have, if, ignore, ignores, in, include, instance, interrupt, into, is, left, less, link, links, migrate, migrates, must, named, not, of, of_type, on, only, opens, or, period, process, properties, property, puts-the-focus, read, reads, receive, receives, referred, reject, rejects, replace, replaces, request, restore, restores, retrieve, retrieves, right, rule, satisfied, select, selects, send, sends, forward, export, pass, sequence, set, sets, shall, split, splits, standard, start, starting, starts, stop, stops, store, stores, synchronize, synchronizes, 
     than, that, the, then, through, ticks, to, type, unselect, unselects, until, update, updates, upload, uploads, use, uses, using, validate, validates, value, values, when, where, while, with, ANY_OTHER, ID, INT, ML_COMMENT, SL_COMMENT, STRING, WS
  },
   keywordstyle = {\color{xtext-keyword}},
   stringstyle=\color{javadocblue},
   commentstyle=\color{codegreen},
   numberstyle=\tiny\color{codegray},
  comment=[l]{\#},
  morestring=[b]',
  morestring=[b]",
  captionpos=b, 
  basicstyle=\scriptsize\ttfamily,
  prebreak=\raisebox{0ex}[0ex][0ex]{\ensuremath{\hookleftarrow}},
}

%% file: introduction.tex
\section{Introduction}
\label{sec:intro}

Requirements typically drive software development and are generally expressed using natural language (NL), which is widely used in many industrial contexts. 
The central role of NL in software requirement specifications (SRSs) stems from its usability in all application domains and its ease of understanding by all stakeholders in software development projects~\cite{PohlKlaus2010Re:f}. A study~\cite{KassabNL14} reports that 61\% of users prefer to express requirements using NL.
However, despite its popularity, NL is highly prone to quality problems, such as vagueness, ambiguity, complexity, and incompleteness~\cite{mavin2010big, FernandezWKFMVC17}. 

One important cause of project failures in industry is quality problems in NL requirements~\cite{ahonen2010software, elizabeth2011requirements}. When such problems are not fixed early during development, they carry over to subsequent development phases, and fixing them becomes a costly and time-consuming process. Improving the quality of NL requirements by identifying quality problems at early development stages is therefore a pivotal need for successful software development. 

In collaboration with an industrial partner, which provides post-trade services for various types of financial securities and develops information systems to support these services, we noticed elevated costs associated with in-house processes to improve the quality of NL requirements, which typically involve several manual inspections and are thus prone to errors. A tool that automatically detects quality problems in NL requirements and guides analysts to improve the quality of NL requirements is thus highly desirable.

Various approaches have been proposed to improve the quality of NL requirements by detecting semantic and syntactic problems, that are often referred as ``smells''~\cite{GenovaFMHM13,FemmerFJKZZ14,FemmerFWE17,FerrariGRTBFG18,EzziniA0SB21}.
For example, \citet{FemmerFWE17} introduced Smella, an automated tool for detecting requirement smells such as ambiguous adverbs and vague pronouns. Smella relies on part-of-speech (POS) tagging, dictionaries, and lemmatization.
However, these approaches do not provide recommendations to analysts on how to rewrite requirements in a disciplined manner to improve their quality. 
Furthermore, existing work~\cite{FemmerFWE17,FerrariGRTBFG18}, which detects a set of quality problems in a requirement, still requires further research to account for many of the recurrent problems faced by analysts. For example, analysts sometimes describe multiple functions in a single requirement (i.e., non-atomic requirement), miss essential words (e.g., actors and verbs) or even phrases (e.g., system responses), or write a requirement following an ambiguous structure (e.g., a system response between conditions). 
We note that some of these quality problems have been studied individually in many prior works, such as checking the completeness~\cite{KaiyaS05,FerraridSG14,EckhardtVFM16,DalpiazSL18,AroraSB19} or ambiguity~\cite{KiyavitskayaZMB08,YangRGWN11,OsamaZAGI20,EzziniA0SB21,EzziniA0S22} of requirements.
However, compared to these research strands, there has been relatively less focus on developing an automated solution that can both detect and resolve multiple quality problems in a requirement. Such a solution is needed to provide a complete picture of the overall requirement quality and thus enable the proposal of a solution that properly fixes all relevant problems together.

To address the challenges stated above, we developed an approach and tool that addresses common quality problems in NL requirements.
In particular, our work aims at assisting analysts with automatically detecting quality problems and suggesting recommendations to fix them in functional requirements.
Given our focus on information systems, a functional requirement specifies what system response an actor is expected to receive when providing certain inputs, if certain conditions are met~\cite{VeizagaATSB21}. 
We have named our tool \TOOL, which means ``solution'' in Quechua. In this article, the term ``smells'' refers to these quality problems in functional requirements that can lead to defects at different levels of severity.
\revision{For example, while the use of passive voice alone in an NL requirement is not necessarily a direct cause of defects, it can contribute to communication issues and misunderstandings, which can in turn increase the risk of defects.
To illustrate this point, we present a real-world example of an NL requirement provided by our industrial partner in the financial domain, anonymized for confidentiality:
\textit{``When an order cancellation message is received then the System\_A GUI  must display the field Reason\_of\_Cancellation''}.
The condition of this NL requirement \textit{``When an order cancellation message is received''} is in passive voice. 
In this requirement, it is difficult for stakeholders to identify the actor receiving the \textit{``order cancellation message''}, which can result in defects in the final product.}
In a functional requirement, a smell indicates a problem located in a specific word or segment (i.e., set of words) that quality analysts should inspect~\cite{FemmerFWE17}.
\TOOL detects nine smells that are commonly present in functional requirements we analyzed in the financial domain, though they are not in any way specific to this domain. Most importantly, \TOOL also provides suggestions for fixing smells and thus improving requirement quality.

\revision{
\textbf{Contributions.} To automatically detect quality problems in functional requirements and provide concrete improvement recommendations, our work relies on a state-of-the-art controlled natural language (CNL) called  Rimay~\cite{VeizagaATSB21}, which is used to specify functional requirements for information systems.
Rimay requires analysts to write requirements in a disciplined manner using controlled grammar and vocabulary, ensuring they follow best practices for writing requirements.
This results in writing precise, unambiguous, complete, and atomic requirements, which in turn enables automated analysis such as traceability analysis~\cite{VeizagaATSBP20} and acceptance criteria generation~\cite{VeizagaATSBP20}.
However, defining a language like Rimay is a different problem than applying it to detect quality problems in non-compliant requirements and propose recommendations for their rectification.
We first define requirement smells corresponding to violations of quality attributes, such as completeness, clarity, atomicity, and correctness, which Rimay enforces when writing requirements.
Based on the Rimay grammar, we define a set of Rimay patterns that capture recommended structures for expressing requirements.
In order to help analysts address smells in a requirement, \TOOL then provides Rimay patterns as suggestions.
Analysts can then rewrite these requirements based on the recommended patterns. Indeed, such patterns indicate what requirement segments are missing, warrant change, or must be re-ordered.
\revision{To automate the detection of requirement smells and the recommendation of Rimay patterns, \TOOL combines natural language processing (NLP) techniques, including tokenization, lemmatization, POS tagging, constituency parsing, glossary search, and Tregex~\cite{LevyA06}.}
We note that this article not only presents an automated method for identifying a large spectrum of smells but, different from existing work, also provides automated and accurate guidance on how to address them.
}


Our empirical study is guided by the following research questions (RQs):
\begin{enumerate}[leftmargin=0em]
    \item[] \textbf{RQ1)~What are the NL requirement smells commonly found in the financial domain?}
    We answer RQ1 by identifying nine smells after analyzing a set of 1404 actual requirements in the financial domain. The smells indicate quality problems in requirements regarding completeness, clarity, atomicity, and correctness. \TOOL is able to detect quality problems in both individual segments of the requirement (segment level) and in the requirement as a whole (requirement level).
    \item[] \textbf{RQ2)~How can smells be detected?} 
    We answer RQ2 by proposing an automated approach that relies on NLP methods to detect the smells in NL requirements. It combines Tregex~\cite{LevyA06} patterns, structural patterns, rules, and glossary search. 
    \item[] \textbf{RQ3)~How can we suggest recommendations to improve requirement quality?}
    We answer RQ3 by proposing an automated approach that suggests a suitable Rimay pattern as a recommendation for an NL requirement with smells. \TOOL first analyzes the overall syntax of an NL requirement. According to this syntax, \TOOL matches and suggests a suitable Rimay pattern. Following our suggested patterns increases the chance of fixing smells. 
    \item[] \textbf{RQ4)~Can \TOOL accurately indicate the occurrence of smells?}
    To answer RQ4, we evaluate \TOOL by measuring the accuracy of identifying smells. To accomplish this, we conducted a case study using 13 SRSs from financial applications. We compared our results against a ground truth established manually by four annotators. Our evaluation results suggest that \TOOL accurately detects smells with a precision and a recall of 89\%. 
    \item[] \textbf{RQ5)~How accurate is \TOOL in recommending requirement patterns to fix smells?}
    RQ5 assesses how well \TOOL suggests appropriate Rimay patterns to fix smells in NL requirements. We compared the performance of \TOOL against a ground truth matching requirements with smells and Rimay patterns, which were manually annotated by four annotators. The results show that \TOOL accurately suggests appropriate Rimay patterns with a precision of 96\% and a recall of 94\%.
\end{enumerate}

To summarize, the main contributions of this work are
(1)~an automated approach that detects smells in SRSs and suggests Rimay patterns to fix these smells and 
(2)~an extensive industrial case study on smell detection, involving a large set of 2725 information system requirements from 13 projects in the financial domain, to assess the accuracy of \TOOL when detecting smells and suggesting patterns to fix them.
Our industrial case study is the largest to date regarding requirements quality assurance.

\textbf{Organization.} The article is structured as follows: Section~\ref{sec:rimay_background} provides an overview of Rimay. Section~\ref{sec:smells and patterns} describes how we derive smells and Rimay patterns. In Section~\ref{sec:qa_approach}, we describe \TOOL in detail. Section~\ref{sec:qa_evaluation} evaluates \TOOL through an industrial case study. Section~\ref{sec:discussion} examines some issues that impact the performance of \TOOL. Sections~\ref{sec:qa_treats} and \ref{sec:qa_related} discuss threats to the validity of our results and related work. Finally, Section~\ref{sec:qa_conclusions} concludes this article.





%% file: background.tex
\section{Controlled Natural Language: Rimay}
\label{sec:rimay_background}

In this section, we introduce Rimay as our work relies on it to define smells and patterns (Section~\ref{sec:smells and patterns}), as well as to identify smells and recommend patterns (Section~\ref{sec:qa_approach}).
Rimay is a controlled natural language (CNL) for writing functional requirements in the domain of information systems~\cite{VeizagaATSB21}. Rimay applies restrictions on vocabulary, grammar, and semantics of NL to allow analysts to write complete, unambiguous, and precise requirements.
Rimay's main grammar rules are inspired by the Easy Approach to Requirements Syntax (EARS) templates~\cite{mavin2009easy}. Many practitioners consider EARS as a good trade-off between flexibility and precision, due to EARS's relatively low training overhead and the quality and readability of the resultant requirements~\cite{MavinWGU16}.
However, EARS templates, which consist of predefined sentence structures with general, coarse-grained concepts and constructs, are not amenable to the type of analyses enabling task automation because they allow the introduction of unstructured and ambiguous text. In contrast to EARS, Rimay provides sentence structures with more specialized, precise, fine-grained concepts and constructs, enabling automated analysis such as reconciliation support between requirement text and models~\cite{VeizagaATSBP20} and automated acceptance criteria generation~\cite{VeizagaATSBP20}.

\begin{figure}
\noindent\begin{minipage}[t]{\linewidth}
\begin{lstlisting}[frame=bt,
label={listing:RequirementRimay},
language=OwnExamples,
linerange=1-1,
breaklines=true,
caption={Overall syntax of a requirement in Rimay.}
]
REQUIREMENT: SCOPE? CONDITION_STRUCTURES? ARTICLE? ACTOR MODAL_VERB not?  SYSTEM_RESPONSE.
\end{lstlisting}
\end{minipage}
\end{figure}

Listing~\ref{listing:RequirementRimay} shows the rule \texttt{REQUIREMENT} specifying the overall syntax of a requirement in Rimay. The rule indicates that the presence of \texttt{SCOPE} and \texttt{CONDITION\_STRUCTURES} is optional, whereas the presence of \texttt{ACTOR}, \texttt{MODAL\_VERB} and \texttt{SYSTEM\_RESPONSE} is mandatory in all requirements.
In a functional requirement, an actor is expected to achieve a system response under certain conditions. 
An actor is a role played by an entity that interacts with the system by exchanging signals, data, or information~\cite{UML25}. 
Furthermore, requirements written in Rimay may have a scope to delimit the effects of the system response. 
For simplicity, in this study, we use ``system response'' to refer to a phrase consisting of an actor, a modal verb, and a system response.




\begin{figure}
\noindent\begin{minipage}[t]{\linewidth}
\begin{lstlisting}[frame=bt,
numbers=left,
xleftmargin=2em,
framexleftmargin=2em,
label={listing:other_rulesRimay},
language=OwnExamples,
linerange=2-13,
breaklines=true,
caption={Syntax of scope and condition structures in Rimay.}
]
SCOPE: For MODIFIER? TEXT (and MODIFIER? TEXT)?,
CONDITION_STRUCTURE: WHILE_STRUCTURE|WHEN_STRUCTURE|WHERE_STRUCTURE|IF_STRUCTURE|TEMPORAL_STRUCTURE
WHILE_STRUCTURE: While PRECONDITION_STRUCTURE
WHEN_STRUCTURE: When TRIGGER
WHERE_STRUCTURE: Where TEXT
IF_STRUCTURE: If PRECONDITION_STRUCTURE|TRIGGER
TEMPORAL_STRUCTURE:  (Before|After|Every) TEXT
CONDITION_STRUCTURES: CONDITION_STRUCTURE ( (,|and|or|,or|,and) CONDITION_STRUCTURE)*, then?
\end{lstlisting}
\end{minipage}
\end{figure}

The Rimay grammar enables analysts to write a wide variety of functional requirements while ensuring that they follow recommended syntactic structures~\cite{VeizagaATSB21}.
For example, Listing~\ref{listing:other_rulesRimay} depicts the grammar rules for scope and condition structures.
As shown in the listing, Rimay's grammar has some common constructs such as \texttt{MODIFIER} (line~1). The construct \texttt{MODIFIER} includes articles (e.g., ``a'', ``an'', and ``the'') and quantifiers (e.g., ``each'', ``all'', ``none'', ``only one'', and ``any''). 
The rule \texttt{SCOPE} (line~1) uses the keyword \CNL{For}, along with the rules \texttt{MODIFIER} and \texttt{TEXT}.

The rule \texttt{CONDITION\_STRUCTURE} (line~2 of Listing~\ref{listing:other_rulesRimay}) defines different ways to use system states, trigger events, and features to express conditions. In a functional requirement, such conditions must hold for the system response to be triggered. 
Furthermore, listing~\ref{listing:other_rulesRimay} (lines 3-7) shows the rules for the following condition structures: \texttt{WHILE}, \texttt{WHEN}, \texttt{WHERE}, \texttt{IF}, and \texttt{TEMPORAL} structures.
The \texttt{WHILE\_STRUCTURE} is used for system responses that are triggered while the system is in one or more specific states. The \texttt{WHEN\_STRUCTURE} is used when a specific triggering event is detected in the system. The \texttt{WHERE\_STRUCTURE} is used for system responses that are triggered only when the system includes particular features. 
These features are described in free form using the rule \texttt{TEXT}.
While Rimay provides fine-grained constructs, it still includes the rule \texttt{TEXT} to handle situations where the use of free text is necessary or desirable.
The \texttt{IF\_STRUCTURE} is used when a specific triggering event happens, or when the system is in a particular state before triggering any system responses. The \texttt{TEMPORAL\_STRUCTURE} is used when the system responses are triggered before or after an event.
Line~8 of Listing~\ref{listing:other_rulesRimay} shows the rule \texttt{CONDITION\_STRUCTURES} that enables the creation of a condition composed of two or more of the conditions mentioned above (lines 3-7).

\begin{table*}[t]
	\centering
	\caption{\revision{Examples of when, temporal, and if conditions, as well as a system response written in Rimay.}}
	\label{tbl:rimay examples}
	\begin{tabularx}{\textwidth}{l X}
		\toprule
		\textbf{Rimay syntax}&\textbf{Rimay example}\\
		\midrule
		\texttt{WHEN\_STRUCTURE} & \tinyCNL$When System-B receives an "email alert" from System-A$ \\
		\arrayrulecolor{lightgray}\hline\arrayrulecolor{black}
		\texttt{TEMPORAL\_STRUCTURE} & \tinyCNL$Before System-A sends an "Instruction" to System-B$ \\
		\arrayrulecolor{lightgray}\hline\arrayrulecolor{black}
		\texttt{IF\_STRUCTURE} & \tinyCNL$If an "Instruction" contains a "Keyword"$ \\
		\arrayrulecolor{lightgray}\hline\arrayrulecolor{black}
		\texttt{SYSTEM\_RESPONSE} & \tinyCNL$The User must upload the "excel file" to System-A$ \\
		\bottomrule
	\end{tabularx}
\end{table*}

\revision{
Table~\ref{tbl:rimay examples} lists Rimay examples of requirement conditions, i.e.,  \texttt{WHEN\_STRUCTURE}, \texttt{TEMPORAL\_STRUCTURE}, and \texttt{IF\_STRUCTURE}, as well as \texttt{SYSTEM\_RESPONSE}.
Note that these examples are independent from one another.
The \texttt{WHEN\_STRUCTURE} and \texttt{TEMPORAL\_STRUCTURE} examples capture the events that trigger system responses, whereas the \texttt{IF\_STRUCTURE} example specifies a precondition required for system responses.
The \texttt{SYSTEM\_RESPONSE} example specifies the action a user must take.
}

\revision{
Thanks to its restrictions on vocabulary, grammar, and semantics of NL, Rimay allows analysts to write functional requirements that satisfy the following quality attributes: \emph{completeness}, \emph{clarity}, \emph{atomicity}, and \emph{correctness}.
\revision{\emph{Completeness} refers to the inclusion of all the information required for the requirement to be complete.
Rimay achieves completeness by having mandatory constructs that ensure the presence of certain contents.
Note that any omission of these constructs will result in syntax errors in a Rimay requirement, making it syntactically incomplete.
For example, Rimay does not allow analysts to write a requirement without a system response.
Such a requirement is thus syntactically incomplete in Rimay.}
\emph{Clarity} refers to the usage of structures, phrases, and words that are free of ambiguity. Rimay achieves clarity by providing a set of predefined structures and restricted vocabulary.
\emph{Atomicity} refers to ensuring that an NL requirement describes a single system function. Rimay enforces that a requirement does not have more than one system response.
\emph{Correctness} refers to the proper use of Rimay's syntax, i.e., the correct arrangements of words and phrases. For example, Rimay does not allow the use of modal verbs in conditions.
}

\revision{
We note that these quality attributes, particularly completeness and clarity, align with those prominently studied in the literature~\cite{MontgomeryFBSM22}, indicating they correspond to commonly observed quality issues in requirements engineering.
Specifically, the completeness of a functional requirement in Rimay (thereafter a Rimay requirement) is based on prior studies~\cite{FemmerFWE17,FerrariGRTBFG18} on the internal completeness~\cite{Zowghi2003OnTI} of a requirement, which concerns the self-containment of a requirement.
The clarity of a Rimay requirement is related to research strands~\cite{OsamaZAGI20,EzziniA0S22} on analyzing requirement ambiguities.
We further discuss these research contributions in Section~\ref{sec:qa_related}.
Regarding the atomicity of a Rimay requirement, it contributes to improving the other quality attributes, such as clarity and verifiability.
For example, a Rimay requirement, precisely describing a single system function, allows for the automatic generation of acceptance criteria for the system function~\cite{VeizagaATSB21, VeizagaATSBP20}.
This, in turn, enhances the traceability between a functional requirement in Rimay and its associated acceptance criteria.
However, the correctness of a Rimay requirement, in this article, is different from the notion of correctness in existing works~\cite{Denger2005}, which concerns whether a requirement accurately captures users' needs.
In our context, as described earlier, the correctness of a Rimay requirement pertains to the correct arrangements of words and phrases defined in the Rimay rules.
Such correct arrangements of words and phrases enable analysts to write requirements in a consistent and unambiguous manner.
}

Rimay is implemented as an add-on editor for Sparx Enterprise Architect~\cite{EA} with the following features: (a)~syntax highlighting to color requirements and format them with different visual styles according to the elements of Rimay, (b)~error marking to automatically highlight the parts of the requirements indicating errors, and (c)~content assisting to automatically, or on request, provide suggestions to analysts on how to complete the statement based on the grammar rules. 

%% file: context.tex
\section{\revision{Industrial Partner and Data Collection}}
\label{sec:partner and data}

\revision{
We conducted this study in collaboration with an industrial partner, who provided the actual requirements for this study and gave feedback while developing \TOOL.
Below, we provide more details about the context of our study and the dataset obtained from our industrial partner.
}

\subsection{\revision{Industrial Partner}}
\label{subsec:partner}

\revision{
Our industrial partner is a leading financial company that provides post-trade services for various types of financial securities and develops information systems to support these services, serving 2500 customers in 110 countries.
Specifically, our collaboration was with their specialized financial services division.
This division is responsible for several tasks, including the development of new financial applications, the maintenance of  existing ones, and the enhancement of applications using advanced technologies.
Their goal is to provide clients with cutting-edge solutions while ensuring compliance with current regulations.
The teams within the division cover project management, service operations, development, testing, and requirements analysis.
}

\revision{
The company employs a methodology rooted in best practices and years of experience for the tasks mentioned above.
For example, financial analysts use natural language (NL) combined with Unified Modeling Language (UML) models to specify requirements.
Their textual NL requirements, written in English, aim to adhere to the Rupp template~\cite{Pohl11}.
The company follows a carefully planned software development process grounded in the V-Model~\cite{sommerville2011}, tailored for stringently regulated industries such as finance.
}

\revision{
We note that the first author was already familiar with the development process, including requirements engineering, of the company, before starting this work, through participating in training sessions and attending numerous meetings.
In addition, all the authors of this article interacted with the industrial partner during the project period through regular bi-weekly meetings and additional meetings as needed to obtain feedback on our progress in defining the problem and developing \TOOL.
For the meetings, the number of industry participants varied between one and five, depending on their availability and interests.
They collectively have over 50 years of industry experience with significant expertise in business analysis, functional design, project management, and requirements engineering.
}

\revision{
The research team of this article collaborated with the company for approximately 1.5 years. 
Our collaboration covered defining the problem this work addresses, identifying smells in requirements, defining Rimay patterns, and developing an automated approach for smell detection and pattern recommendation.
Hence, our work was motivated by practitioners' needs and was validated in a practical context.
However, given its scale, the evaluation of the \TOOL prototype was completed after the project ended, including data annotation, implementation, and evaluation.
Since \TOOL is an automated analysis tool, which does not require human intervention, we were able to successfully evaluate \TOOL without significant involvement from our industrial partner. 
}

\subsection{\revision{Data Collection}}
\label{subsec:data}

\revision{
To conduct this study, we first collected actual NL requirements from our industrial partner.
Financial analysts provided us with a set of 13 representative SRSs, written by different analysts and containing various numbers of requirements.
These SRSs describe various types of projects, including functional updates to existing applications, making existing applications compliant with new regulations, creating new applications from scratch, and the migration of existing applications to new platforms.
We refer to these SRSs as SRS1-SRS13, belonging to set $S$.
They contain 2725 requirements in total, which makes this industrial case study the largest to date regarding requirements quality assurance. 
}

\begin{table}[t]
	\centering
	\caption{\revision{Distribution of actual requirements received from our industrial partner.}}
	\label{table:dataset}
	\begin{tabularx}{\columnwidth}{l Y Y Y}
		\toprule
		\textbf{Subset} & \textbf{SRS ID} & \# \textbf{Requirements} & \# \textbf{Words} \\
		\midrule
		\multirow{6}{*}{$S^D$}
			& SRS1 & 192 & 6363 \\
			&  SRS2 & 188 & 7462 \\ 
			& SRS3 & 118 & 5699 \\
			& SRS4 & 161 & 7925 \\
			& SRS5 & 451 &  22057 \\
			& SRS6 & 294 & 14573 \\
		
		\arrayrulecolor{lightgray}\hline\arrayrulecolor{black}
		\multirow{7}{*}{$S^E$}
			& SRS7 & 367 & 9512 \\
			& SRS8 & 90 & 3573  \\
			& SRS9 & 167 & 6211 \\
			& SRS10 & 192 & 8892 \\
			& SRS11 & 19 & 331 \\
			& SRS12 & 340 & 15448 \\
			& SRS13 & 146 & 7539 \\            
		
		\midrule
		\midrule
		\textbf{Total} & & 2725 & 115585 \\
		\bottomrule
	\end{tabularx}
\end{table}

\revision{
Table~\ref{table:dataset} shows the distribution of requirements, along with the total word count, across SRSs.
We randomly split the set $S$ of 2725 requirements into two subsets $S^D$ and $S^E$, containing 1404 and 1321 requirements, respectively.
The subset $S^D$ contains SRSs SRS1-SRS6 and was used for developing \TOOL (see Sections~\ref{sec:smells and patterns} and \ref{sec:qa_approach}).
The subset $S^E$ contains SRSs SRS7-SRS13 and was used for evaluating \TOOL (see Section~\ref{sec:qa_evaluation}).
}

%% file: input.tex
\section{Requirements Smells and Rimay Patterns}
\label{sec:smells and patterns}

This section describes the process we conducted to derive requirements smells and patterns from NL requirements and the Rimay language. We define a catalog of smells that identify the syntactic and semantic errors commonly found in NL requirements.
In addition, we introduce Rimay patterns that describe predefined structures for high-quality requirements expressed in Rimay. These patterns will be used as suggestions for analysts to rewrite requirements containing smells, thereby improving the quality of the requirements.

\subsection{Requirements Smells}
\label{subsec:smells}

This section aims to answer RQ1:~\textbf{What are the NL requirement smells commonly found in the financial domain}? 
\revision{For RQ1, we inspected a set $S^D$ of 1404 NL requirements, which were written by different analysts across six financial systems, as described in Section~\ref{subsec:data}.
\revision{In order to review the requirements in a consistent manner, we used the quality attributes enforced by Rimay as our review criteria.
These criteria enabled us to identify requirements that have quality problems with respect to completeness, clarity, atomicity, and correctness.
In this article, the term ``requirement smell'' refers to these quality problems in functional requirements that may lead to misunderstandings, which can, in turn, increase the risk of defects in the product.}}
\revision{We note that there are existing collections of requirement smells reported in prior work~\cite{FemmerFWE17,FerrariGRTBFG18}, which are further discussed in Section~\ref{sec:qa_related}.
Drawing upon these works, we cross-referenced our findings with these existing smell collection.
In our context, for the reasons mentioned above, we apply Rimay to support a practical solution to detect a large variety of common smells and rectify them.
Hence, we chose to inspect the 1404 NL requirements using the quality attributes enforced by Rimay.}


\begin{table*}[t]
\centering
\caption{Catalog of nine smells. The examples in this table are derived from the actual requirements used in our case study but are anonymized for confidentiality.}
\label{table:reqSmells}
\begin{tabularx}{\textwidth}{>{\hsize=0.2\hsize}X X >{\hsize=0.2\hsize}X}
\toprule
\textbf{Smell name} & \textbf{(D) Description and (E) Example} & \textbf{Quality Attribute}\\
\midrule
Non-atomic\newline requirement & (D)~Non-atomic requirement refers to a requirement that has more than one action in the system response. (E)~\textit{``System-A must add System-B to their downstream systems and allow System-C to subscribe to the Reporting flow.''}& Atomicity \\
\arrayrulecolor{lightgray}\hline\arrayrulecolor{black}
Incomplete\newline requirement & (D)~Incomplete requirement refers to a requirement that does not have a system response but has other optional segments, i.e., condition and scope. (E)~\textit{``When System-A receives a message from Security Manager and if the message is part of the B-file, according to the mapping rules.''} & Completeness\\
\arrayrulecolor{lightgray}\hline\arrayrulecolor{black}
Incorrect order\newline requirement & (D)~Incorrect order requirement refers to a requirement that its condition segment is located after its system response. This arrangement of requirement segments may lead to a vague interpretation of the condition. (E)~\textit{``When the user is on the Utilities page and the user clicks on the button ``Display on main page'', System-A must open the Alert section when the user launches System-A.''} & Correctness\\
\arrayrulecolor{lightgray}\hline\arrayrulecolor{black}
Coordination\newline ambiguity & (D)~Coordination ambiguity refers to a requirement that has two or more conditions and these conditions are connected by a coordinated conjunction ``or''. (E)~\textit{``When System-A performs eligibility check for a participant or if the holding type is complex or if the holding type is simple and the F-value is Prime, then System-B must ...''} & Clarity\\
\arrayrulecolor{lightgray}\hline\arrayrulecolor{black}
Not requirement & (D)~Not requirement refers to a statement that does not contain any requirement segment, i.e., scope, condition, and system response. (E)~\textit{``The R6 instruction defines the original instruction.''} & Correctness\\
\arrayrulecolor{lightgray}\hline\arrayrulecolor{black}
Incomplete\newline condition & (D)~Incomplete condition indicates a condition that lacks either actor or verb. (E)~\textit{``Upon receipt of a message in the message Queue, System-A must set the state to unprocessed.''} & Completeness\\
\arrayrulecolor{lightgray}\hline\arrayrulecolor{black}
Incomplete\newline system response & (D)~Incomplete system response refers to a system response that lacks either an actor, a modal verb, or a verb. (E)~\textit{``When the user clicks on the Filter button, System-A opens the Filter screen.''} & Completeness\\
\arrayrulecolor{lightgray}\hline\arrayrulecolor{black}
Passive voice & (D)~Passive voice indicates that the condition or system response in a requirement is described in the passive voice. Such requirements likely miss actors. (E)~\textit{``When a rejection order is received for a cancellation request, System-A must raise a web alert.''} & Completeness\\
\arrayrulecolor{lightgray}\hline\arrayrulecolor{black}
Not precise verb & (D)~Not precise verb refers to a verb used in the condition or system response in a requirement that is not precise enough. Such a verb does not define a precise action. The list of our not precise verbs includes: ``accomplish'', ``account'',  ``come'', ``consider'', ``default'', ``define'', ``do'', ``get'', ``make'', ``perform'', ``process'', ``propose'', ``raise'', ``read'', ``support'', and ``want''. This list is curated by the analysts at our industrial partner. (E)~\textit{``System-A must be able to process System-B's instructions with input media INPUT.''}  & Clarity\\
\bottomrule
\end{tabularx}
\end{table*}

\revision{
In our inspection process, the first author of this article manually reviewed the requirements to uncover a list of those that violate the quality attributes and to define a catalog of smells.
\revision{We note that this author has extensive expertise and practical experience in Rimay from defining the language to (re)writing requirements in it, as presented in the author's previous work~\cite{VeizagaATSB21}.
Given this background, the author was ideally suited to review the requirements and pinpoint violations of the quality attributes that Rimay enforces.}
We randomly divided the set $S^D$ of 1404 requirements into six distinct subsets for incremental inspection.
Specifically, $S^D_1$ contains 35\% of $S^D$, and each $S^D_i$ contains 13\% of $S^D$, where $i \in \{2,\ldots,6\}$.
We then sequentially inspected each subset of requirements in the order $S^D_1$, $S^D_2$, $\ldots$, $S^D_6$.
During each inspection, we identified requirements that violate the quality attributes and characterized these violations as requirement smells.
For example, if an inspected requirement had multiple actions in its system response, we identified it as violating atomicity and characterized the violation as the non-atomic requirement smell.
\revision{After inspecting each subset of requirements, we obtained feedback from our industrial partner to validate that the catalog of smells we defined aligns with common errors frequently made by analysts when writing NL requirements.
In addition to such feedback, the other authors closely monitored progress through regular meetings.}
The catalog of smells was thus continuously refined and expanded throughout the inspection process.
}

\revision{
We applied the notion of saturation~\cite{GlaserSaturat} to determine when to stop the inspection process.
Saturation occurs when no new information can be gained from the data being analyzed.
We continued the inspection process as long as we detected any instances of new smells that were not already listed in our smell catalog.
From this process, we defined a catalog of nine smells: non-atomic requirement, incomplete requirement, incorrect order requirement, coordination ambiguity, not requirement, incomplete condition, incomplete system response, passive voice, and not precise verb. 
Table~\ref{table:reqSmells} lists the nine smells. 
The first column shows the smell names, and the second column provides the smell descriptions and examples. 
We note that the examples listed in this table are derived from the actual requirements used in our case study. 
However, they are sanitized for confidentiality. 
The third column indicates the quality attribute that each smell violates. What is noteworthy from the nine smells we identified is that they seem rather generic and probably applicable to information systems in other domains.
}


\begin{tcolorbox}[enhanced jigsaw,left=2pt,right=2pt,top=0pt,bottom=0pt,opacityback=0,colframe=black,coltext=black]
\emph{The answer to RQ1 is that}, based on our inspection, the following NL requirement smells are commonly found in the financial domain: non-atomic requirement, incomplete requirement, incorrect order requirement, coordination ambiguity, not requirement, incomplete condition, incomplete system response, passive voice, and not precise verb.
These smells violate atomicity, completeness, correctness, and clarity of NL requirements. None of these smells appears to be specific to the financial domain.
\end{tcolorbox}

\subsection{Rimay Patterns}
\label{subsec:rimay_patterns}
This section describes the process conducted to derive requirements patterns from the Rimay language. These patterns describe predefined structures of Rimay concepts. As we describe in Section~\ref{sec:rimay_background}, Rimay provides specialized concepts and constructs to specify functional requirements. However, Rimay does not provide patterns that guide analysts on how to rewrite a requirement with smells.

\begin{figure*}[t]
\centering
 \centerline{\includegraphics[width=1\linewidth]{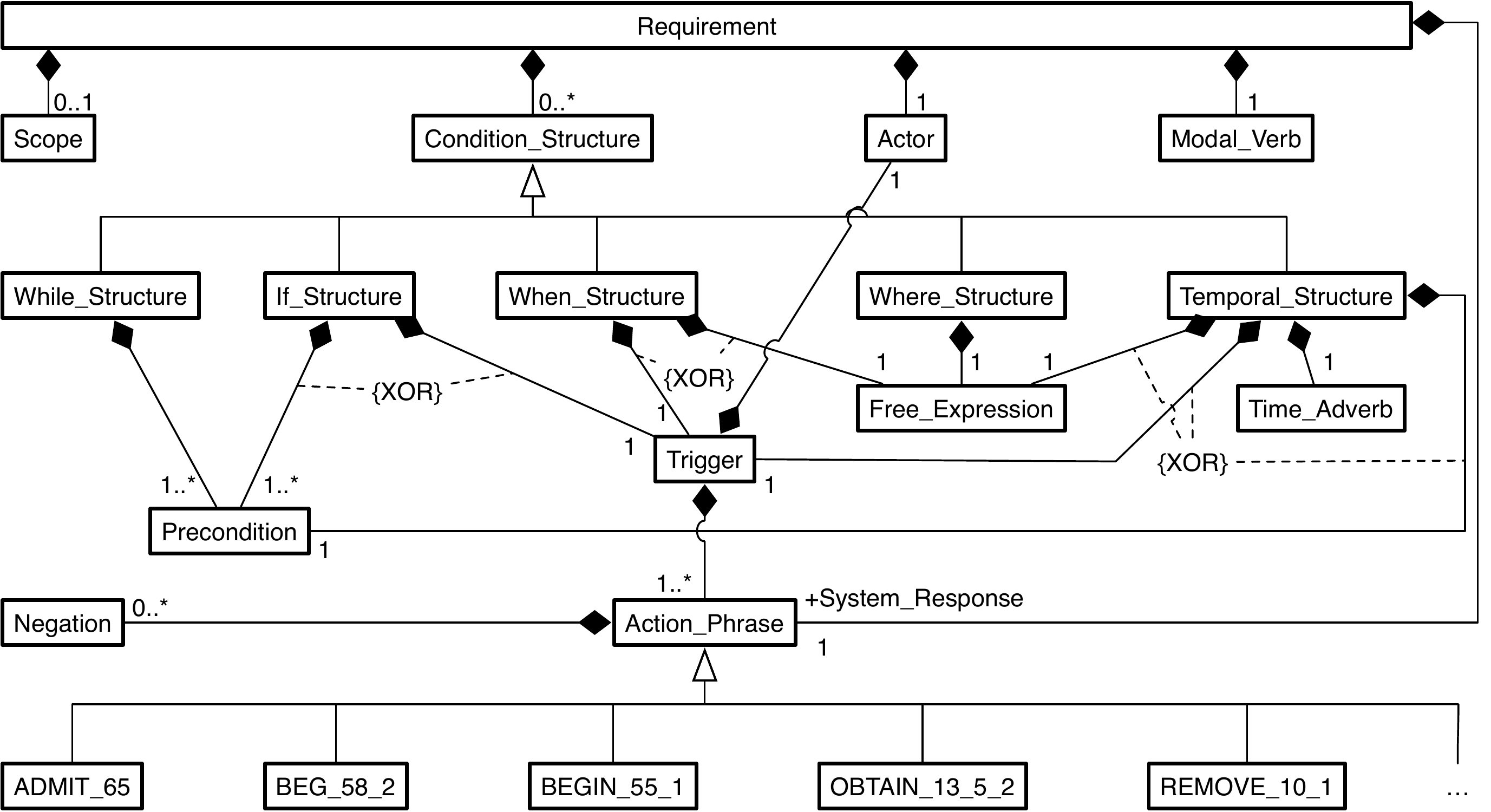}}
\caption{Rimay conceptual model.}
\label{fig:rimayConceptualModel}
\end{figure*}

To derive the Rimay patterns, we first created a conceptual model for capturing the concepts and their relations underlying the Rimay language. Figure~\ref{fig:rimayConceptualModel} shows the conceptual model of Rimay. The model captures five concepts to define ``Requirement' as follows: ``Scope'', ``Condition\_Structure'', ``Actor'', ``Modal\_Verb'', and ``Action\_Phrase''. ``Condition\_Structure'' is further specialized into five concepts: ``While\_Structure'', ``If\_Structure'', ``When\_Structure'', ``Where\_Structure'', and ``Temporal\_Structure''. 
Rimay restricts the ways in which these conditions can be expressed. For example, ``While\_Structure'' can be used for system responses that are triggered while the system is in a particular state (see the relation between ``While\_Structure'' and ``Precondition'' in Figure~\ref{fig:rimayConceptualModel}).
The concept ``Action\_Phrase'' is specialized into 58 concepts (e.g., ``ADMIT\_65'' and ``BEG\_58\_2'') that correspond to the grammar rule names defined in Rimay~\cite{VeizagaATSB21}. 
Note that the grammar rules of the ``Action\_Phrase'' concept describe the syntactic structure and vocabulary allowed.
Figure~\ref{fig:rimayConceptualModel} shows only a few concepts for ``Action\_Phrase''. 

\begin{table*}[t]
	\centering
	\caption{\revision{Concepts of action phrases (shown in Figure~\ref{fig:rimayConceptualModel}) and their Rimay examples (excerpted from the previous work on Rimay~\cite{VeizagaATSB21}).}}
	\label{tbl:action phrase examples}
	\begin{tabularx}{\textwidth}{l X}
		\toprule
		\textbf{Concept}&\textbf{Example: Rimay Action Phrase}  \\
		\midrule
		ADMIT\_65 & 
		\tinyCNL$exclude the "Gregorian dates that are not business days" in the System based on "the relevant calendar"$  \\
		\arrayrulecolor{lightgray}\hline\arrayrulecolor{black}
		BEG\_58\_2 & \tinyCNL$request the System to "cancel the settlement" by using the "Order Reference"$  \\
		\arrayrulecolor{lightgray}\hline\arrayrulecolor{black}
		BEGIN\_55\_1 & \tinyCNL$start the "calculation of the next NAV date on daily basis"$  \\
		\arrayrulecolor{lightgray}\hline\arrayrulecolor{black}
		OBTAIN\_13\_5\_2 &
		\tinyCNL$receive a DA_file from CFCL_IT$  \\
		\arrayrulecolor{lightgray}\hline\arrayrulecolor{black}
		REMOVE\_10\_1 & \tinyCNL$delete the "DECU field" from the "Settlement Parties block"$  \\
		\bottomrule
	\end{tabularx}
\end{table*}

\revision{
Table~\ref{tbl:action phrase examples} presents examples of Rimay action phrases corresponding to the concepts that specialize the ``Action\_Phrase'' concept in Figure~\ref{fig:rimayConceptualModel}.
For example, ``BEG\_58\_2'' restricts the usage of the word ``request'' in expressing an action phrase.
Specifically, the Rimay syntax of ``BEG\_58\_2'' restricts that ``request'' can be followed by an optional modifier, a mandatory actor, an optional ``for'' or ``to'' phrase, and an optional ``by using'' phrase, in this specific order.
The example  \CNL$request the System to "cancel the settlement" by using the "Order Reference"$  follows the Rimay syntax.
We note that the exact Rimay syntax for action phrases, along with examples of Rimay action phrases, is provided in the previous work on Rimay~\cite{VeizagaATSB21}.
}

\revision{
To identify Rimay patterns, we utilized both the Rimay conceptual model (Figure~\ref{fig:rimayConceptualModel}) and grammar.
We traversed the syntax tree of the grammar to derive possible combinations of Rimay concepts.
In Rimay, a functional requirement is composed of scope, condition, and system response segments.
These segments correspond to the ``Scope'', ``Condition\_Structure'', and ``Action\_Phrase'' concepts, respectively, as shown in Figure~\ref{fig:rimayConceptualModel}.
According to the Rimay grammar (see Listing~\ref{listing:RequirementRimay}), the scope, condition, and system response segments of a functional requirement  must appear in that specific order.
In addition, Rimay specifies that the scope and condition segments are optional in a Rimay requirement, while the system response segment is mandatory.
Regarding the condition segment in a Rimay requirement, analysts can express the following three condition concepts: ``Precondition'', ``Trigger'', and ``Temporal\_Structure''.
These concepts respectively correspond to a condition in which the requirement can be invoked (``Precondition''), an event that initiates the requirement (``Trigger''), and a temporal event that occurs either before or after the requirement's invocation (``Time\_Adverb'').
Furthermore, Rimay allows a condition segment to specify multiple conditions.
Hence, in Rimay, analysts have the flexibility to specify a condition segment as either a precondition, a trigger condition, a time condition, or a combination of these, referred to as multiple conditions.
Since Rimay provides analysts with 2 options for scope (i.e., inclusion and exclusion), 5 options for condition (i.e., inclusion --- precondition, trigger condition, time condition, and multiple conditions --- and exclusion), and 1 option for system response (i.e., inclusion), a Rimay requirement can be classified into 10 distinct patterns, as shown in Table~\ref{fig:rimayPatterns}.
}

\begin{table*}[t]
\centering
\caption{Rimay patterns}
\label{fig:rimayPatterns}
\begin{tabularx}{\textwidth}{>{\hsize=0.5\hsize}X X >{\hsize=0.5\hsize}X }
\toprule
\textbf{Pattern Name}&\textbf{Rimay Pattern}&\textbf{Mapping to Conceptual Model} \\
\midrule
1. Scope and system response & 
\tinyCNL$For each|all|... "Text" ,|then the? Actor must <Action> (every "Text")?.$ & Scope, Actor, Modal\_Verb,\newline and Action\_Phrase\\
\arrayrulecolor{lightgray}\hline\arrayrulecolor{black}
2. Scope, condition (precondition), and system response & \tinyCNL$For each|all|... "Text", if <Property> is equal to | is less or equal to |... "Value" ,|then the? Actor must <Action> (every "Text")?.$ & Scope, Precondition, Actor,\newline Modal\_Verb, and Action\_Phrase\\
\arrayrulecolor{lightgray}\hline\arrayrulecolor{black}
3. Scope, condition (trigger), and system response & \tinyCNL$For each|all|... "Text", when the? Actor <Action> (every "Frequency")?,|then the? Actor must <Action> (every "Text")?.$ & Scope, Trigger, Actor,\newline Modal\_Verb, and Action\_Phrase\\
\arrayrulecolor{lightgray}\hline\arrayrulecolor{black}
4. Scope, condition (time), and\newline system response &
\tinyCNL$For each|all|... "Text", after|before "Text" ,|then the? Actor must <Action> (every "Text")?.$ & Scope, Time\_Adverb, Actor,\newline Modal\_Verb, and Action\_Phrase \\
\arrayrulecolor{lightgray}\hline\arrayrulecolor{black}
5. System response & \tinyCNL$The? Actor must <Action> (every "Text")?.$ & Actor, Modal\_Verb,\newline and Action\_Phrase \\
\arrayrulecolor{lightgray}\hline\arrayrulecolor{black}
6. Condition (precondition) and system response & \tinyCNL$If <Property> is equal to | is less or equal to |... "Value" ,|then the? Actor must <Action> (every "Text")?.$ & Precondition, Actor, Modal\_Verb,\newline and Action\_Phrase \\
\arrayrulecolor{lightgray}\hline\arrayrulecolor{black}
7. Condition~(trigger) and\newline system response & \tinyCNL$When the? Actor <Action> (every "Frequency")? ,|then the? Actor must <Action> (every "Text")?.$ & Trigger, Actor, Modal\_Verb,\newline and Action\_Phrase \\
\arrayrulecolor{lightgray}\hline\arrayrulecolor{black}
8. Condition~(time) and\newline system response & \tinyCNL$After|Before "Text" ,|then the? Actor must <Action> (every "Text")?.$ & Time\_Adverb, Actor,\newline Modal\_Verb, and Action\_Phrase\\
\arrayrulecolor{lightgray}\hline\arrayrulecolor{black}
9. Scope, multiple conditions, and\newline system response & \tinyCNL$For each|all|... "Text", if <Property> is equal to | is less or equal to |... "Value" ,|and  when the? Actor <Action> (every "Frequency")? ,|then the? Actor must <Action> (every "Text")?.$ & Scope, Condition Structure (two or more), Actor, Modal\_Verb,\newline and Action\_Phrase \\ 
\arrayrulecolor{lightgray}\hline\arrayrulecolor{black}
10. Multiple conditions and\newline system response & \tinyCNL$If <Property> is equal to | is less or$
\newline \tinyCNL$equal to |... "Value" ,|and  when the? Actor <Action> (every "Frequency"),|then the? Actor must <Action> (every "Text")?.$ & Condition Structure (two or more), Actor,  Modal\_Verb,\newline and Action\_Phrase \\ 
\bottomrule
\end{tabularx}
\end{table*}

Table~\ref{fig:rimayPatterns} outlines the 10 Rimay patterns derived by combining the concepts captured in the conceptual model of Rimay. The first column shows the pattern names. The second column specifies the pattern in the Rimay syntax. Finally, the third column indicates the combination of the Rimay concepts used to derive the pattern. Table~\ref{fig:rimayPatterns} does not include all the keywords defined in Rimay and does not include the templates for Action\_Phrases. We refer interested readers to our previous work on Rimay~\cite{VeizagaATSB21} for a complete reference to the concepts and constructs of the Rimay language.

\revision{
\revision{The derived patterns will be used by \TOOL to provide suggestions to analysts when \TOOL detects any smells (see Table~\ref{table:reqSmells}) in NL requirements.
The patterns, as guidance, are intended to help analysts rewrite the NL requirements in Rimay to improve their quality.
Indeed, such
patterns indicate what requirement segments are missing,
warrant change, or must be re-ordered.
Recall that Rimay allows analysts to write requirements that satisfy the following quality attributes: completeness, clarity, correctness, and atomicity.
Since \TOOL detects requirement smells that violate these quality attributes, proposing appropriate Rimay patterns will assist analysts in addressing the identified smells in NL requirements.}
We note, however, that \TOOL is not designed to automatically rewrite NL requirements in Rimay to rectify any smells.
Such automatic translation is probably not possible in general as it would require automated additions of missing segments or changes, which we leave out of the scope of this study.
}




%% file: approach.tex
\section{Approach}
\label{sec:qa_approach}

\begin{figure*}[t]
\centering
 \centerline{\includegraphics[width=0.95\linewidth]{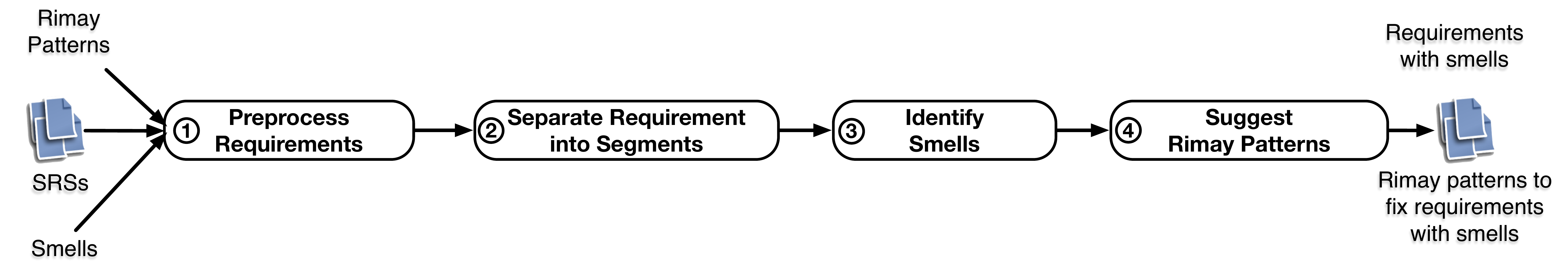}}
\caption{Approach overview.}
\label{fig:approachOverview}
\end{figure*}

Figure~\ref{fig:approachOverview} provides an overview of the four steps of \TOOL. The inputs are (1)~software requirements specifications (SRSs) that contain a set of functional requirements, (2)~Rimay patterns~(Section~\ref{subsec:rimay_patterns}), and (3)~the catalog of nine smells commonly found in NL requirements (Section~\ref{subsec:smells}). 

\begin{figure*}[!t]
\centering
 \centerline{\includegraphics[width=1.00\linewidth]{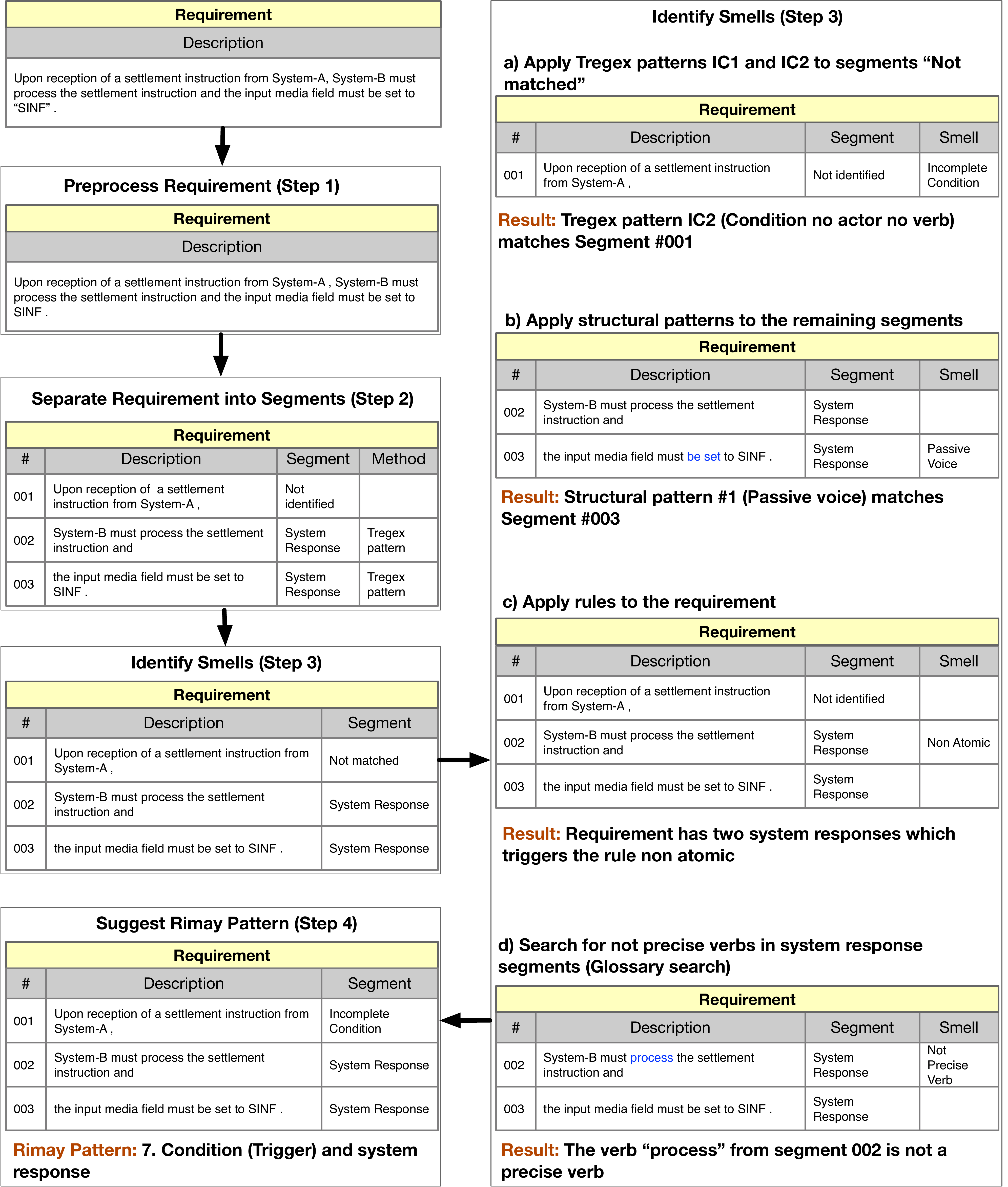}}
\caption{Examples illustrating smell detection and Rimay pattern suggestions.}
\label{fig:approachExample}
\end{figure*}

As shown in Figure~\ref{fig:approachOverview}, in Step~1, \TOOL applies preprocessing steps to the NL requirements extracted from the SRSs. In Step~2, requirements are separated into segments (e.g., scope, condition, and system response), relying on patterns and a segment splitter. The patterns are written using Tregex, which is a language for defining patterns in text syntax trees.  
In Step~3, \TOOL detects smells in NL requirements by applying several techniques, such as Tregex patterns, structural patterns, rules, and glossary search. In our context, a structural pattern refers to a pattern that checks the sequence of words in a requirement. 
Finally, in Step~4, \TOOL suggests a pattern for analysts to fix the NL requirements with smells and convert them into Rimay. Rimay helps reduce the risk of having quality problems in requirements since it has precise syntax and semantics. 
In this section, we describe in detail all the steps of \TOOL using the running example shown in Figure~\ref{fig:approachExample}.

\subsection{Step 1: Preprocess Requirements}
\label{subsec:qa_preprocess}

We apply a set of preprocessing steps to the NL requirements extracted from SRSs, including tokenization (dividing the text of the requirement into tokens, such as punctuation marks and words), post-tagging (assigning part of speech tags to tokens, such as pronouns, verbs, and adjectives), and constituency parsing (a process that identifies the structural units of sentences, e.g., clause, noun phrase, verb phrase).
We also remove single and double quotes but keep the metadata in requirements (i.e., the MS Word metadata in our case study). Such metadata includes line breaks and bullet points and are, in our context, useful for detecting multiple conditions and system responses.
Figure~\ref{fig:approachExample} (Step 1) shows an example of a preprocessed requirement. \TOOL removes the double quotes since we observe that single and double quotation marks prevent us from correctly identifying the structural units of the sentences when using a constituency parsing algorithm (e.g., AllenNLP~\cite{AllenNLP2017}).  


\subsection{Step 2: Separate Requirement into Segments}
\label{subsec:qa_separate_req}

This step automatically separates an NL requirement into segments (e.g., scope, condition, and system response) to (1)~analyze each segment of the requirement independently with the purpose of finding smells (Step~3 of Figure~\ref{fig:approachOverview}) and (2)~determine the overall syntax of the requirements with the purpose of suggesting a precise Rimay pattern (Step~4 of Figure~\ref{fig:approachOverview}). 
We create an automated procedure to separate a requirement into segments using the following methods: (1)~Tregex patterns and (2)~segment splitters.

\textbf{Tregex patterns.} 
The Tregex query language allows users to define regular expression-like patterns in tree structures~\cite{LevyA06}. Tregex is designed to define patterns that involve the tree nodes and the hierarchical relations among the tree nodes in the syntax tree of a requirement.  
To separate a requirement into segments, we created a set of patterns using Tregex.
In our context, a Tregex pattern matches a specific structure of the constituency structure of a requirement. The constituency structures of the requirements are obtained in Step~1 (Section~\ref{subsec:qa_preprocess}).


To create Tregex patterns, we analyzed the syntax of 1404 requirements that are identical to the requirements used in Section~\ref{subsec:smells}. 
The process to derive the patterns is as follows: 
(1)~We group the requirements that have the same segments. A segment could be a scope, condition, and system response. (2)~We analyze the constituency structures of the segments of each group. (3)~We derive Tregex patterns that match the constituency structure of the segments of each group.
(4)~We analyze all patterns matching the same segment across all groups, attempting to refine and merge them, if possible. If not, a segment may have more than one pattern. 



\begin{table*}[t]
\centering
\caption{\revision{Tregex patterns to identify segments in requirements. The examples in this table are derived from the actual requirements used in our case study but are anonymized for confidentiality.}}
\label{tab:TregPatterns}
\begin{tabularx}{\textwidth}{l>{\hsize=0.17\hsize}X X X}
\toprule
\textbf{ID}  & \textbf{Segment} &  \textbf{Tregex Pattern} &  \textbf{Example}\\
\midrule
SC1 & Scope    & (\texttt{PP} \texttt{<} ((\texttt{IN} \texttt{<} For) \texttt{\$+} \texttt{NP})) \texttt{>} (\texttt{S} \texttt{<} (/(\texttt{RB} \texttt{|} , \texttt{|} \texttt{ADVP})/ \texttt{\$+} (\texttt{NP} \texttt{\$+} \texttt{VP}) )) & \emph{For all the depositories}, System-A must create a T30 transaction processing command.\\
\arrayrulecolor{lightgray}\hline\arrayrulecolor{black}

SC2 & Scope    & ( (\texttt{PP} \texttt{<} ((\texttt{IN} \texttt{<} For) \texttt{\$+} \texttt{NP})) [ \texttt{>-} ((\texttt{VP} \texttt{<} \texttt{MD}) \$- \texttt{NP}) ] \texttt{\&!}\texttt{>}\texttt{>} \texttt{PP}) & System-A must create an instruction with the the remote code value ``B'' \emph{for each settlement request.}\\
\arrayrulecolor{lightgray}\hline\arrayrulecolor{black}
C1	& Condition  & \texttt{WHADVP} \$+ (\texttt{S} \texttt{<}, (\texttt{S} \texttt{<} (\texttt{NP} \$+ \texttt{VP})))) \texttt{>} \texttt{SBAR} \&\texttt{!>>} \texttt{VP} & \emph{When System-A creates one of the following reports: list, list with Beta}, System-A must populate the field ``A'' in the report. \\
\arrayrulecolor{lightgray}\hline\arrayrulecolor{black}
C2	& Condition     & ((\texttt{IN} \texttt{<} \texttt{once})  \$+ (\texttt{S} \texttt{<} (\texttt{NP} \$+ \texttt{VP}))) [\texttt{>} \texttt{SBAR} \texttt{|} \texttt{>} \texttt{S}] \& \texttt{>} !\texttt{VP} & \emph{Once System-A has successfully validated a settlement request}, System-A must send an acknowledge message to System-B.\\
\arrayrulecolor{lightgray}\hline\arrayrulecolor{black}
C3	& Condition     & \texttt{SBAR} \texttt{<} ( \texttt{WHADVP}  \$+ \texttt{S} \texttt{<} (\texttt{NP} \$++ \texttt{VP})) & System-A must send a settlement request to System-B \emph{when the contract note has been received from System-C.}\\
\arrayrulecolor{lightgray}\hline\arrayrulecolor{black}
C4	& Condition     & (\texttt{WHADVP} \$+ (\texttt{S} \texttt{<} (\texttt{NP} \$+ \texttt{VP})) !\texttt{>>} /(\texttt{VP} \texttt{|} \texttt{SBAR})/) & \emph{When the fund frequency in reference data has an empty value}, then System-A must set the fund frequency to ``daily''.\\
\arrayrulecolor{lightgray}\hline\arrayrulecolor{black}

C5	& Condition     & (\texttt{WHADVP} !\texttt{<} /(\texttt{of} \texttt{|} \texttt{to})/)  \$+ (\texttt{NP} \$+ \texttt{VP}) !\texttt{>>} /(\texttt{VP} \texttt{|} \texttt{SBAR})/ & \emph{When the user clicks on the left side menu, portfolio section}, System-A must display the portfolios.\\
\arrayrulecolor{lightgray}\hline\arrayrulecolor{black}
C6	& Condition     & (\texttt{SBAR} \texttt{<} ((\texttt{WHADVP} !\texttt{<<} \texttt{that}) \$+ (\texttt{S} \texttt{<<} (\texttt{SBAR} \texttt{<<}, /(\texttt{That} \texttt{|} \texttt{that})/)  \& \texttt{<<}, (\texttt{NP} \$++ \texttt{VP} \texttt{|} \$++ \texttt{VB})))) !\texttt{>>} \texttt{VP} & \emph{When a user confirms that he wants to cancel the creation of an account record}, the System-A must  delete the related parameters recorded in account with status ``draft''.\\
\arrayrulecolor{lightgray}\hline\arrayrulecolor{black}
C7	& Condition     & (\texttt{SBAR} \texttt{<} ((\texttt{WHADVP} !\texttt{<<} \texttt{that}) \$+ (\texttt{S} !\texttt{<<} \texttt{SBAR} \&\texttt{<<}, (\texttt{NP} \$++ \texttt{VP} \texttt{|} \$++ \texttt{VB})))) !\texttt{>>} \texttt{VP}  & \emph{If the settlement date is present in the instruction sent to System-A} then System-A must store in data storage unit.\\
\arrayrulecolor{lightgray}\hline\arrayrulecolor{black}
C8	& Condition Time & (\texttt{PP} \texttt{<} (\texttt{IN} \texttt{<} (/\^ (\texttt{after} \texttt{|} \texttt{before})\$/) \$+ (\texttt{NP} !\texttt{<} \texttt{VB} \texttt{<} \texttt{NN}) )) \texttt{>} \texttt{S} & \emph{Before the System-A cutover}, System-A must update the entity-A data model. \\
\arrayrulecolor{lightgray}\hline\arrayrulecolor{black}
SR1 & System Response & \texttt{NP} \$+ (\texttt{VP} \texttt{<} ( \texttt{MD} ?\$+ \texttt{ADVP} \$++ (\texttt{VP} \texttt{<<}, (/\texttt{VB}.?/  \$+ (\texttt{S} \texttt{<} (\texttt{NP} \$++ \texttt{VP})))))) \texttt{>} \texttt{S}  & \emph{System-A must send the fund details report to local team daily.}\\


\bottomrule
\end{tabularx}
\caption*{\color{black}\texttt{S}:~Clause, \texttt{SBAR}:~Subordinate clause, \texttt{WHADVP}:~Wh-adverb phrase, \texttt{VP}:~Verb phrase, \texttt{VBG}:~Verb gerund, \texttt{NP}:~Noun phrase, \texttt{PP}:~Prepositional phrase, \texttt{IN}:~Preposition, \texttt{NN}:~Noun, \texttt{RB}:~Adverb, \texttt{ADVP}:Adverb phrase, \texttt{MD}:~Modal, and \texttt{VB}:~Verb}
\end{table*}

Table~\ref{tab:TregPatterns} presents the 11 Tregex patterns that we derived after following the process described above. In total, we derived two patterns that detect scope (see SC1 and SC2 in Table~\ref{tab:TregPatterns}), eight patterns that detect conditions (C1–C8), and one pattern that detects system response (SR1).
\revision{In the table, for each Tregex pattern, there is an example requirement containing the segment matched by the pattern. The segment is emphasized in italic.}

\begin{figure*}[t]
\centering
 \centerline{\includegraphics[width=0.9\linewidth]{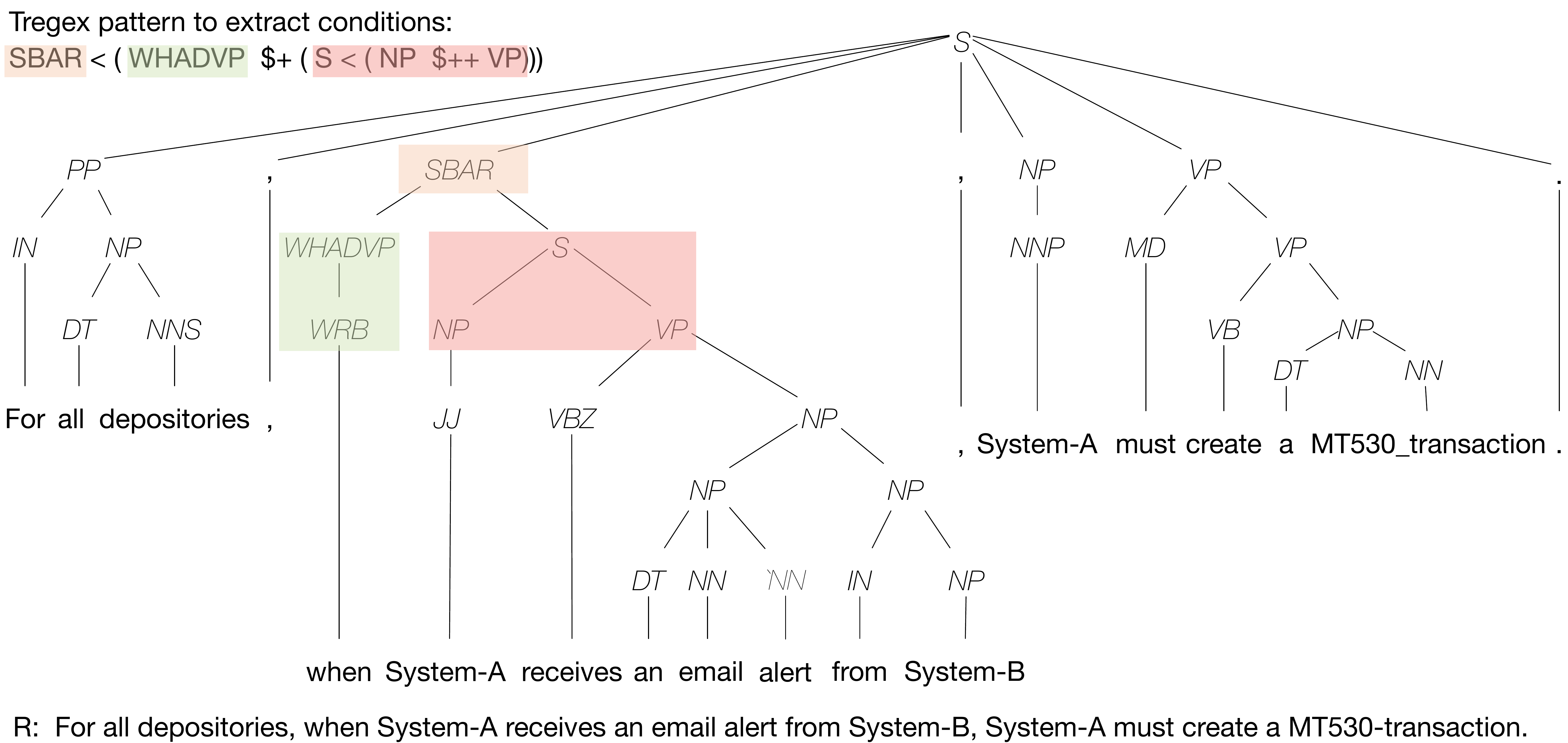}}
\caption{An example Tregex pattern that matches a constituency parsing tree of a condition.}
\label{fig:TregexExample}
\end{figure*}

Figure~\ref{fig:TregexExample} shows an example of the usage of the pattern C3 (see Table~\ref{tab:TregPatterns}) to extract the condition of requirement R from the corresponding constituency parsing tree. The requirement has a scope (``For all depositories''), a condition (``when System-A receives an email alert from System-B''), and a system response (``System-A must create an MT530\_transaction''). We note that some concepts of requirement R are anonymized to comply with the confidentiality agreement with our industrial partner.
The condition is composed of a subordinate conjunction (\texttt{WHADVP}:~``when''), a noun phrase (\texttt{NP}:~``System-A''), and a verb phrase (\texttt{VP}:~``receives an email alert from System-B'').  
As shown in Figure~\ref{fig:TregexExample} (highlighted in the colored boxes), the pattern C3 identifies a subordinate clause (\texttt{SBAR}) that is the parent ($<$) of a Wh-adverb phrase (\texttt{WHADVP}), which is the immediate left sister (\$+) of a clause (\texttt{S}). Clause (\texttt{S}) is the parent ($<$) of a noun phrase (\texttt{NP}), which is the left sister (\$++) of a verb phrase~(\texttt{VP}).

\textbf{Segment splitter.} 
We propose a segment splitter that attempts to separate requirements into segments (e.g., scope, condition, and system response). The segment splitter is only used when all the Tregex patterns fail to identify segments in a requirement. 
We note that such failures can be caused by a malformed constituency parsing tree that incorrectly identifies the structural units of a requirement or a requirement structure that was not observed during the creation of the Tregex patterns.

To create our segment splitter, we first collect the keywords that characterize the beginning of each segment in a requirement. These keywords include (1)~for the condition:  ``when'', ``if'', ``where'', ``while'', and ``until'', (2)~for the scope: ``for'', and (3)~for the system response: ``then'', (line break), ``;'', ``else'', and ``otherwise''. We identified these keywords based on both the Rimay grammar and the requirements inspected when deriving our catalog of requirements smells (see Section~\ref{subsec:smells}).
In general, such an analysis method that relies on keywords has limitations in terms of exhaustiveness. When needed, however, extending our set of keywords is very straightforward.

Our segment splitter splits the requirement into segments by detecting the above keywords in the requirements and then validates each segment. A segment is considered valid if it has the mandatory information content (e.g., actor, modal verb, and verb in a system response). Table~\ref{tab:info_content} shows the mandatory information content that each segment should have to be considered valid.

\begin{table*}[t]
\centering
\caption{Information contents characterizing the requirement segments}
\label{tab:info_content}
\begin{tabularx}{\textwidth}{ >{\hsize=0.2\hsize}X X}

\toprule
\textbf{Segment} & \textbf{Information Content} \\
\midrule
Scope & \mbox{\exCNL{For [each | all | none] noun}} \\
\arrayrulecolor{lightgray}\hline\arrayrulecolor{black}
Condition & \mbox{\exCNL{[When | if | where | while | until] noun verb}} \\        
\arrayrulecolor{lightgray}\hline\arrayrulecolor{black}
System response & \mbox{\exCNL{[then | <line break> | ; | else | otherwise] noun modal-verb verb}} \\ 
\bottomrule
\end{tabularx}

\end{table*}

Finally, when segments do not comply with their mandatory information contents, \TOOL labels them as ``Not Matched''. These segments will be further analyzed to detect smells in Step~3 (Section~\ref{subsec:qa_detect_smells}).

In the example shown in Figure~\ref{fig:approachExample} (Step~2), we first apply all the Tregex patterns (Table~\ref{tab:TregPatterns}) to the requirement. The pattern SR1 is matched, and segments 002 and 003 are identified as system responses. \TOOL then applies the segment splitter. However, the segment splitter fails to identify segment 001, since the word ``Upon'' is not present in our keyword list.
Since this word describes a preposition, it can be anywhere in a requirement and not only at the beginning of a segment.

\subsection{Step 3: Identify Smells}
\label{subsec:qa_detect_smells}

\begin{algorithm}[t]
	\caption{\revision{Detect Smells in NL Requirement.}}
	\label{alg:identify smells}
	\begin{algorithmic}[1]
		\Input
		\Statex $r$: NL requirement
		\Output
		\Statex $D$: detected requirement smells
		\Statex
		
		\State $\mathit{SEG} \gets \Call{SeparateSegments}{r}$  \textcolor{gray}{// Step 2}
		\State $D \gets \{\}$
		\ForEach{$\mathit{seg}$ $\in$ $\mathit{SEG}$}
		\State $D_t \gets \Call{ApplyTregexPattern}{\mathit{seg}}$
		\State $D_s \gets \Call{ApplyStructuralPattern}{\mathit{seg}}$
		\State $D_r \gets \Call{ApplyRules}{\mathit{seg}}$
		\State $D_g \gets \Call{SearchGlossary}{\mathit{seg}}$
		\State $D \gets D \cup D_t \cup D_s \cup D_r \cup D_g$
		\EndFor
		\State \Return $D$
	\end{algorithmic}
\end{algorithm}

This section answers \textbf{RQ2: How can smells be detected?} 
\revision{Algorithm~\ref{alg:identify smells} presents our automated procedure for detecting any of the smells introduced in Section~\ref{subsec:smells}.
As shown in line~1 of the algorithm, we analyze the requirement segments obtained from Step~2 in Figure~\ref{fig:approachOverview}.
To detect smells in each segment, the procedure employs the following techniques: Tregex patterns, structural patterns, rules, and glossary search (see lines 3-9).
Below, we describe in detail these techniques and the smells that each of them detects.}

\textbf{Tregex patterns.}
Recall from Section~\ref{subsec:qa_separate_req} that the Tregex patterns match the specific structures of the constituency parsing tree resulting from an NL requirement. While analyzing the set of requirements in Section~\ref{subsec:qa_separate_req}, we observed two groups of requirements that contain incomplete conditions.

The following are two examples of these requirements: EIC1 and EIC2.
\textit{``EIC1: Upon reception from System-A the status Pending of an Instruction, then ...''}. 
The condition of EIC1 misses the actor and the verb. Instead of a verb, the condition contains the noun ``reception'', making it unclear who or what is doing the receiving.
\textit{``EIC2: When creating a new participant, System-A must...''}. The condition of EIC2 lacks the actor, and the verb is described using a gerund.
Missing an actor or a verb in a condition statement can result in ambiguity and incompleteness issues in a functional requirement.
To detect such conditions, we derived two Tregex patterns. To do so, we found a set of 55 requirements from the 1404 requirements (i.e., $S^D$ in Section~\ref{subsec:smells}) that are similar to EIC1 and EIC2. 
Next, we grouped the requirements into two sets. Each set shares the same information content. Then, for each set, we derived a Tregex pattern. Table~\ref{tab:tregex_incomplete} shows the two derived patterns, i.e., IC1 and IC2, to detect incomplete conditions.

\begin{table*}[t]
\centering
\caption{Tregex pattern to detect incomplete conditions}
\label{tab:tregex_incomplete}
\begin{tabularx}{\textwidth}{>{\hsize=0.1\hsize}X X}
\toprule
\textbf{ID} & \textbf{Tregex Pattern}  \\
\midrule
 IC1 & ((SBAR $<$ (WHADVP \$+ (S $<$ ((VP $<$ (VBG \$+ NP | \$+ PP)) !\$++ NP !\$-- NP)))) !$>>$ VP ) \\
 \arrayrulecolor{lightgray}\hline\arrayrulecolor{black}
 IC2 & ((PP $<$ ( (IN $<$ Upon) \$+ (NP $<$ ( (NP $<<$ NN) \$++ PP )))) !$>>$ /(VP | SBAR)/) \\
\bottomrule
\end{tabularx}
\caption*{\texttt{S}:~Clause, \texttt{SBAR}:~Subordinate clause, \texttt{WHADVP}:~Wh-adverb phrase, \texttt{VP}:~Verb phrase, \texttt{VBG}:~Verb gerund, \texttt{NP}:~Noun phrase, \texttt{PP}:~Prepositional phrase, \texttt{IN}:~Preposition, and  \texttt{NN}:~Noun }

\end{table*}

In the example shown in Figure~\ref{fig:approachExample} (Step~3a), \TOOL matches the pattern IC2 with segment 001 ``Upon reception of a settlement instruction from System-A''. The condition lacks the actor and the verb; therefore, the smell ``Incomplete condition'' is detected.

\textbf{Structural patterns.} 
This method analyzes the segments of a requirement using structural patterns to detect the following smells: ``Incomplete condition'', ``Incomplete system response (SR)'', and ``Passive voice''.
Regarding the smell ``Incomplete condition'', recall that our Tregex patterns aim at detecting it. However, due to various reasons, in practice, the constituency parsing tree of a requirement may be malformed. In such cases, our structural patterns are applied to detect the smell. 
A structural pattern checks the presence of certain words describing the concepts (see Figure~\ref{fig:rimayConceptualModel}) of segments in a specific sequence.

As shown in Table~\ref{table:structuralPatterns}, we define 14 structural patterns to detect the above smells. 
We analyzed the 1404 requirements by categorizing them into different smell groups. For each group, we inspected the requirements at the segment level (e.g., condition and system response) based on Rimay's concepts and constructs and identified specific structural patterns to detect the smells.
There are eight patterns that check for the smell ``Passive voice'' in the following tenses: present simple, present perfect, past simple, and past perfect. In addition, we define three patterns that check for the smell ``Incomplete condition'', and three patterns to detect the smell ``Incomplete system response''.

We analyze the segments obtained from Step~2 of Figure~\ref{fig:approachOverview} using the structural patterns. For example, the structural pattern  ``Passive voice \#7'' matches the condition of the following requirement ``When/\texttt{WRB} an/\texttt{DT} Order/\texttt{NNS} has/\texttt{VBZ} been/\texttt{VBN} assigned/\texttt{VBN} via/\texttt{IN} propagation/\texttt{NNP} ..''. The condition contains verb ``has''(v$_{1}$) in the present tense, followed by verb ``been''(v$_{3}$) in the past participle, and followed by verb ``assigned''(v$_{4}$) in the past participle.

In the example shown in Figure~\ref{fig:approachExample} (Step~3b), \TOOL applies our structural patterns to the segments of the requirement. Specifically, the structural pattern ``Passive Voice 1'' matches segment 003. This segment has the verb ``be''(v$_{1}$) in its base form and the verb ``set''($v_{4}$) in its past participle form.

\begin{table*}[t]
\centering
\caption{Structural patterns for smell detection}   
\label{table:structuralPatterns}
\begin{tabularx}{\textwidth}{>{\hsize=0.3\hsize}X  >{\hsize=0.1\hsize}X >{\hsize=0.3\hsize}X X}
\toprule
\textbf{Smell} &  \textbf{\#} & \textbf{Structural Pattern} & \textbf{Example} \\
\midrule
\multirow{8}{*}{Passive voice} & 1 &   \texttt{v$_{1}$}   v$_{4}$ & ``...is taken...'' \\
& 2 & \texttt{v$_{1}$  adv  v$_{4}$}   & ``...has not taken...''\\
& 3 & \texttt{v$_{2}$  v$_{4}$}        & ``...was taken...''  \\
& 4 & \texttt{v$_{2}$  adv  v$_{4}$} & ``...was not taken...''    \\
& 5 & \texttt{v$_{2}$  v$_{3}$  v$_{4}$}  & ``...had been taken...'' \\
& 6 & \texttt{v$_{2}$  adv  v$_{3}$  v$_{4}$}  & ``...had not been taken...'' \\
& 7 & \texttt{v$_{1}$  v$_{3}$  v$_{4}$}  &  ``...has been taken...''\\
& 8 & \texttt{v$_{1}$  adv  v$_{3}$  v$_{4}$} & ``...has not been taken...''\\
\arrayrulecolor{lightgray}\hline\arrayrulecolor{black}
Incomplete condition  & 9 &\texttt{sc  o$_{1}$} & ``When for each subscriptions...'' \\
& 10 & \texttt{sc v$_{1}$} & ``When receives the subscription order ...''  \\
& 11 & \texttt{sc n$_{1}$ o$_{1}$} & ``When the System-A seennd the subscription order ...''  \\
\arrayrulecolor{lightgray}\hline\arrayrulecolor{black}
Incomplete & 12 & \texttt{md v$_{1}$} & ``then must send the settlement request...'' \\
system response& 13& \texttt{n$_{1}$ v$_{1}$} & ``System-A closes the Filter screen...'' \\
& 14 & \texttt{n$_{1}$ md o$_{2}$} & ``System-B must sed the settlement request...''\\ [0.1cm]

\bottomrule
\end{tabularx}
\caption*{\texttt{v$_{1}$}:~Verb base form (be $|$ have), \texttt{v$_{2}$}:~Verb past (be $|$ have), \texttt{v$_{3}$}:~Verb past participle (be), \texttt{v$_{4}$}:~Verb past participle, \texttt{adv}:~Adverb (not), \texttt{sc}:~Subordinated conjunction, \texttt{n$_{1}$}:~Noun, \texttt{md}:~Modal verb, \texttt{o$_{1}$}:~Other word than noun and verb, \texttt{o$_{2}$}:~Other word than verb, \texttt{p}:~Preposition (for), \texttt{q}:~Quantifier, \texttt{o$_{3}$}:~Other word than noun}
\end{table*}

\textbf{Rules.} We propose a set of rules to analyze the segments of a requirement identified by Step~2 of Figure~\ref{fig:approachOverview}, aiming at detecting the following smells: ``Non-atomic'', ``Incomplete requirement'', ``Incorrect order requirement'', ``Coordination ambiguity'', and ``Not a requirement''.
Our rules analyze how the segments are connected to each other and determine the sequence of the segments of the requirement. 
In the following, we describe these rules:
\begin{enumerate}
    \item \textbf{Non-atomic:} The requirement has more than one segment for system response.
    \item \textbf{Incomplete requirement:} The requirement misses the system response segment, but it has other segments, such as scope and condition.
    \item \textbf{Incorrect order requirement:} The requirement has one or more condition segments after its system response.
    \item \textbf{Coordination ambiguity:} When the requirement has two or more subsequent conditions, \TOOL extracts the word(s) or character(s) that separate the conditions. If the separator(s) is  the conjunction \textit{``or''}, then \TOOL triggers the smell ``Coordination ambiguity''.
    \item \textbf{Not a requirement:} All segments of the requirement are neither scope, condition, nor system response.
\end{enumerate}

In the example depicted in Figure~\ref{fig:approachExample} (Step~3c), \TOOL applies our rules to the requirement. The rule ``Non-atomic'' identifies two system responses in the requirement that triggers the smell ``Non-atomic''.

\textbf{Glossary search.}
This method aims to identify the smell: ``Not a precise verb''.
\revision{For this purpose, we create a glossary of verbs that do not describe precise actions, are difficult to test, and carry several meanings.
Note that these criteria are used to define vague words in the literature~\cite{Berry2003}.}
For example, according to the English dictionary, the verb ``process'' means ``operate on (data) by means of a program''. This verb does not provide a precise action, which makes it difficult for an analyst to test the requirements that contain such a verb.
To create our glossary of verbs, we gather all the verbs of the requirements used in Section~\ref{subsec:smells}. We search for verbs that do not have precise actions and are difficult to test. Our glossary includes the following verbs: accomplish, account, base, come, consider, default, define, do, get, make, perform, process, propose, make, raise, read, support, and want. Our glossary also includes verbs that have several meanings but only one etymology (polysemy). These verbs in our glossary are: come and get. 

We have elaborated our glossary in collaboration with two experts working for our industrial partner. The experts agreed that they prefer to avoid using these verbs when specifying requirements as they are not precise enough and are indeed difficult to test.

To detect these verbs in the requirements, \TOOL automatically extracts verbs from the requirement segments condition and system response.
Next, \TOOL obtains the lemmas of these verbs.
Finally, \TOOL searches for the lemma in our glossary. If there is a match, \TOOL triggers the smell ``Not a precise verb''.

In the example shown in Figure~\ref{fig:approachExample} (Step~3d), \TOOL analyzes the segments 002 and 003 because they are system responses; then, \TOOL extracts the verb ``process'' because it belongs to our glossary. Hence, the smell ``Not a precise verb'' is detected.

\begin{tcolorbox}[enhanced jigsaw,left=2pt,right=2pt,top=0pt,bottom=0pt,opacityback=0,colframe=black,coltext=black]
\emph{The answer to RQ2 is that} \TOOL detects smells in NL requirements using several complementary techniques: Tregex patterns, structural patterns, rules, and glossary search. 
\TOOL combines these techniques to ensure that various requirement smells are accurately identified since no single technique is comprehensive enough to detect all of them.
\end{tcolorbox}

\subsection{Step 4. Suggesting Rimay Patterns}
\label{subsec:qa_rimay_patterns}

This section answers \textbf{RQ3: How can we suggest recommendations to improve requirement quality?} To suggest Rimay patterns as recommendations, \TOOL analyzes the segments of a requirement identified by Step~2 to match one of the 10 Rimay patterns (Section~\ref{subsec:rimay_patterns}). 
The matched pattern will guide analysts when fixing any smell detected in a requirement and converting it into a Rimay requirement.

\begin{algorithm}[t]
\caption{\revision{Suggest Rimay Pattern.}}
\label{alg:suggest patterns}
\begin{algorithmic}[1]
\Input
\Statex $r$: NL requirement
\Output
\Statex $p$: suggested Rimay pattern
\Statex

\State $\mathit{SEG} \gets \Call{SeparateSegments}{r}$  \textcolor{gray}{// Step 2}
\State \textcolor{gray}{// Count segment frequencies}
\State $\mathit{cnt}_s \gets \Call{CountScopes}{\mathit{SEG}}$
\State $\mathit{cnt}_c \gets \Call{CountConditions}{\mathit{SEG}}$
\State $\mathit{cnt}_r \gets \Call{CountSystemResponses}{\mathit{SEG}}$
\State \textcolor{gray}{// Identify condition type}
\State $\mathit{ctype} \gets \Call{IdentifyConditionType}{\mathit{SEG}}$
\State \textcolor{gray}{// Match Rimay pattern}
\State $p \gets \Call{MatchRimayPattern}{\mathit{cnt}_s,\mathit{cnt}_c,\mathit{cnt}_r,\mathit{ctype}}$
\State \Return $p$
\end{algorithmic}
\end{algorithm}

\revision{
Algorithm~\ref{alg:suggest patterns} shows \TOOL's automated procedure for suggesting a Rimay pattern for a given NL requirement.
To identify a suitable Rimay pattern, \TOOL first counts the frequencies of the segments in a requirement, since the Rimay patterns are defined in terms of segments (see lines 1-5).
More concretely, \TOOL counts the number of segments---scopes, conditions, and system responses---that appear in a requirement.
For the condition segments, \TOOL further classifies them into trigger, time, and precondition (Section~\ref{sec:rimay_background}) (see lines 6-7). 
To identify ``time condition'', \TOOL checks if a segment is matched by the pattern ``C8''~(Table~\ref{tab:TregPatterns}) in Step~2.
To identify ``precondition condition'', \TOOL extracts the verb phrase (VP) from a condition segment. If the VP contains one of the operators~\cite{VeizagaATSB21}, i.e.,  ``is equal to'', ``less or equal to'', ``contain'', and ``have'', then the condition segment is identified as ``precondition condition''.
Moreover, to identify ``trigger condition'', we extract the VP from a condition segment. If the VP contains verbs other than the ones used by the ``precondition condition'', then the segment is classified as ``trigger condition''. \TOOL also counts the frequencies of incomplete segments with the following smells:  ``incomplete condition'' and ``incomplete system response''.
Once the frequencies are calculated, \TOOL maps the frequencies to any of the 10 Rimay patterns (see lines 8-9). 
Table~\ref{tab:freqSegRimayPat} presents the frequency of segments for each of the 10 Rimay patterns.
The first column shows the name of the pattern. From the second to the sixth column, we show the frequency of each segment contained in each of the 10 Rimay patterns.}

\begin{table*}[t]
	\aboverulesep=0ex
	\belowrulesep=0ex
	\centering
	\caption{Segment frequencies in Rimay patterns.}
	\label{tab:freqSegRimayPat}
	\begin{tabularx}{\textwidth}{X |>{\hsize=0.1\hsize}Y |>{\hsize=0.25\hsize}Y |>{\hsize=0.25\hsize}Y |>{\hsize=0.25\hsize}Y |>{\hsize=0.2\hsize}Y}
		\toprule
		\multicolumn{1}{c|}{\multirow{2}{*}{\textbf{Pattern}}}& \multirow{2}{*}{\textbf{Scope}} & \multicolumn{3}{c|}{\textbf{Condition}} & \multicolumn{1}{c}{\multirow{2}{*}{\textbf{System Response}}} \\
		\cmidrule{3-5}
		
		&& \textbf{Pre-condition} & \textbf{Trigger} & \textbf{Time} & \\
		\midrule
		1. Scope and system response & 1 & 0 & 0 & 0 & 1 \\
		\arrayrulecolor{lightgray}\hline\arrayrulecolor{black}
		2. Scope, condition (precondition), and system response & 1 & 1 & 0 & 0 & 1 \\
		\arrayrulecolor{lightgray}\hline\arrayrulecolor{black}
		3. Scope, condition (trigger), and system response & 1 & 0 & 1 & 0 & 1 \\
		\arrayrulecolor{lightgray}\hline\arrayrulecolor{black}
		4. Scope, condition (time) and system response & 1 & 0 & 0 & 1 & 1 \\
		\arrayrulecolor{lightgray}\hline\arrayrulecolor{black}
		5. System response & 0 & 0 & 0 & 0 & 1 \\
		\arrayrulecolor{lightgray}\hline\arrayrulecolor{black}
		6. Condition(precondition) and system response & 0 & 1 & 0 & 0 & 1 \\
		\arrayrulecolor{lightgray}\hline\arrayrulecolor{black}
		7. Condition (trigger) and system response & 0 & 0 & 1 & 0 & 1 \\
		\arrayrulecolor{lightgray}\hline\arrayrulecolor{black}
		8. Condition (time) and system response & 0 & 0 & 0 & 1 & 1 \\
		\arrayrulecolor{lightgray}\hline\arrayrulecolor{black}
		
		9. Scope, multiple conditions, and system response & 1 & \multicolumn{3}{c|}{2 or more} & 1 \\
		\arrayrulecolor{lightgray}\hline\arrayrulecolor{black}
		10. Multiple conditions and system response & 0 &\multicolumn{3}{c|}{2 or more} & 1 \\
		\bottomrule
	\end{tabularx}
\end{table*}

In the example shown in Figure~\ref{fig:approachExample} (Step~4), \TOOL analyzes the three segments 001, 002, and 003 in the requirement.
Segment 001 is an incomplete condition as \TOOL does not detect any verb in the segment. 
However, the Tregex pattern (i.e., IC2 in Table~\ref{tab:tregex_incomplete}) identifies the smell ``Incomplete condition'', indicating that the condition uses the noun ``reception'' instead of a verb. 
This analysis result suggests that the verb is an action verb, indicating that it is a condition trigger. 
To summarize, the requirement in Figure~\ref{fig:approachExample} (Step~4) has a condition trigger and two system responses.
Since Rimay discourages analysts from writing non-atomic requirements, \TOOL suggests Rimay pattern ``7. Condition(Trigger) and system response'' to help analysts rewrite the requirement.

\begin{figure}[t]
	\centering
	\centerline{\includegraphics[width=\columnwidth]{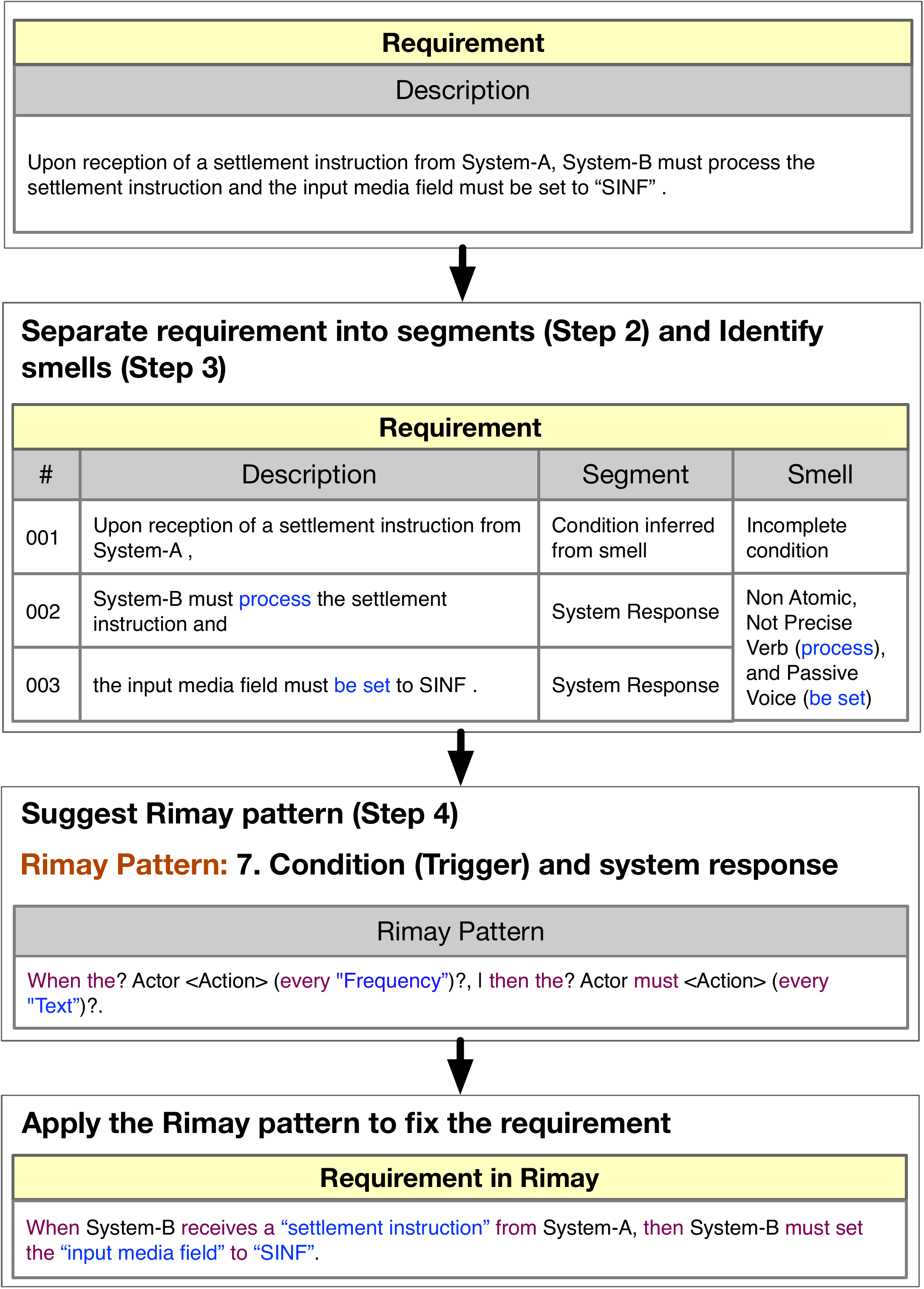}}
	\caption{\revision{Application of a Rimay pattern.}}
	\label{fig:apply pattern}
\end{figure}

\revision{Figure~\ref{fig:apply pattern} illustrates how an analyst applies the suggested Rimay pattern to address smells in an NL requirement.
Given the suggested Rimay pattern ``7. Condition(Trigger) and system response'' (see Table~\ref{fig:rimayPatterns}), the analyst rewrites the NL requirement in Rimay to address the following smells: ``incomplete condition'', ``non-atomic requirement'', ``not precise verb'', and ``passive voice''.
The Rimay requirement includes the trigger condition: \CNL$When System-B receives a "settlement instruction" from System-A$, addressing the ``incomplete condition'' smell, and the system response: \CNL$System-B must set the "input media field" to "SINF"$, addressing the ``non-atomic requirement'' and ``passive voice'' smells.
In this example, we note that the analyst identified segment 002 in the NL requirement as unnecessary and consequently removed it.
}

\begin{tcolorbox}[enhanced jigsaw,left=2pt,right=2pt,top=0pt,bottom=0pt,opacityback=0,colframe=black,coltext=black]
\emph{The answer to RQ3 is that} \TOOL suggests recommendations in the form of Rimay patterns by analyzing segments (i.e., scope, condition, and system response) in an NL requirement.
The recommended Rimay patterns are meant to guide analysts when fixing identified smells in an NL requirement and converting it into a Rimay requirement. In summary, \TOOL indicates segments in the requirements that may be missing, are incorrectly ordered, or that require change.  
\end{tcolorbox}

%% file: evaluation.tex
\section{Evaluation}
\label{sec:qa_evaluation}

In this section, we describe the case study conducted to address RQ4 and RQ5. We follow best practices for reporting case study research in software engineering~\cite{Runeson2012}.

\input{evaluation_case_study_design}

\input{evaluation_data_collection}
\input{evaluation_collecting_evidence}

\input{evaluation_metrics}
\input{evaluation_implementation}
\input{evaluation_analys_data}

%% file: evaluation_case_study_design.tex
\subsection{Case Study Design}
\label{subsec:qa_case_study_design}

Our evaluation aims to answer the following RQs:
\begin{enumerate}[leftmargin=0em]
    \item []\textbf{RQ4.} Can \TOOL accurately indicate the occurrence of smells?    
    \item[]\textbf{RQ5.} How accurate is \TOOL in recommending requirement patterns to fix smells?
\end{enumerate}

To answer RQ4 and RQ5, we measured the accuracy of \TOOL in detecting smells and suggesting Rimay patterns, using an annotated dataset as a ground truth for comparison.
\revision{To construct our ground truth $GT$, we annotated the 13 SRSs (described in Section~\ref{subsec:data}) provided by our industrial partner.
To annotate the SRSs, we had four annotators who manually analyzed the syntax and semantics of each requirement to detect smells and suggest a Rimay pattern.
Sections~\ref{subsec:qa_data_collection} and \ref{subsec:annotated_requirements} provide further details about our annotation process and results.}

\revision{Once the ground truth was completed, we developed \TOOL using the development set $S^D$ of requirements (i.e., SRS1-SRS6), as described in Section~\ref{subsec:qa_implementation}. 
We then evaluated the accuracy of \TOOL in detecting smells and suggesting Rimay patterns.
To this end, we compared the results obtained by \TOOL using the evaluation set $S^E$ (i.e., SRS7-SRS13) against the ground truth $GT$.
We thus calculated the accuracy of smell detection and pattern recommendation using $S^E$, which we report in Section~\ref{sebsec:qa:analysis_results}.}

%% file: evaluation_data_collection.tex
\subsection{\revision{Data Annotation}}
\label{subsec:qa_data_collection}

\revision{To annotate the 13 SRSs described in Section~\ref{subsec:data}, we hired and trained three external annotators to identify smells in SRSs and recommend Rimay patterns to fix such smells.}
Two out of the three external annotators have more than two years of experience in requirements engineering. The other external annotator has knowledge about the domain of our industrial partner and also has more than two years of experience in requirements engineering.
To train the annotators, we randomly selected and used 70 requirements from $S^D$. During the training, we applied Cohen's kappa~\cite{Cohen1960} to measure inter-annotator agreement and obtained a score of 0.89, indicating strong agreement. 

After the training, the three external annotators and the first author of this article conducted a collective 312-hour annotation process.
We note that the annotations made by the first author of this article are included in the subset $S^D$ and hence were used only to develop \TOOL.
We assigned the SRSs to the annotators accounting for their individual situations and contracts, which resulted in different numbers of SRSs being assigned to each annotators. 
The annotators manually analyzed the syntax and semantics of each requirement in the SRSs of $S^D$ and $S^E$ to detect smells and assign a Rimay pattern. 
After the annotators completed annotating each SRS, we analyzed the annotation results by having an in-person monitoring session (30-60 minutes). In each session, the annotators pointed out the requirements that were difficult to annotate. We then discussed them to reach an agreement on the correct annotations.
Once the annotation process concluded, we randomly selected 10\% of the requirements annotated in $S$ for inspection. From the analysis results, we found that most of the annotations (more than 80\%) were satisfactory, indicating agreement among the annotators. For the unsatisfactory annotations (less than 20\%), we then identified their causes and asked the annotators to correct them throughout all SRSs in $S$. We finally accepted the annotations.


%% file: evaluation_collecting_evidence.tex
\subsection{Annotated Requirements}
\label{subsec:annotated_requirements}

Table~\ref{table:smellsDataset} shows the annotation results of the smells detected in $S$. The first column indicates the smell name. The second through seventh columns present the number of requirements containing the smell listed in each row for each batch in $S$. The last column displays the total number of occurrences of each of the nine smells in $S$.
Table~\ref{table:smellsDataset} confirms that the SRSs used in our study are highly diverse in terms of smells since there is significant variation across batches in terms of the distribution of smells.
Further, smells occur according to very different frequencies. We found that ``8. Passive Voice'' is the smell with the highest frequency in $S$  (accounting for 40.9\% of the requirements) while,
in contrast, we have smells such as ``5. Not a requirement'', with very low frequencies (0.28\% of the requirements).

\begin{table*}[t]
\centering
\caption{Smells - Annotation results for set $S$}
\label{table:smellsDataset}
\begin{tabularx}{\textwidth}{X >{\hsize=0.2\hsize}Y >{\hsize=0.2\hsize}Y >{\hsize=0.2\hsize}Y >{\hsize=0.2\hsize}Y >{\hsize=0.2\hsize}Y >{\hsize=0.2\hsize}Y >{\hsize=0.2\hsize}Y}
\toprule
 \textbf{Smell} & \textbf{Batch 1} & \textbf{Batch 2} & \textbf{Batch 3} & \textbf{Batch 4} & \textbf{Batch 5} & \textbf{Batch 6}& \textbf{Total}  \\
\midrule  
1. Non-atomic               & 14 & 79 & 107 & 6 &  23 &  78 & 310 \\
2. Incomplete requirement   & 2 &  8 & 1 & 1 & 1 & 2 & 15\\ 
3. Incorrect order requirement  & 21 & 40 & 34 & 9 & 7 & 14 & 125\\
4. Coordination ambiguity       & 0 & 4 & 0 & 1 & 5 & 4 & 14 \\
5. Not a requirement        & 2 & 0 & 0 & 1 &  1 &  1 & 5\\ 
6. Incomplete condition     & 18 & 72 & 70 & 9 & 49 & 104 & 322\\
7. Incomplete system response    & 4 & 9 & 0 & 2 & 1 & 3 & 19\\
8. Passive voice        & 174 & 219 & 59 & 33 & 95 & 141 & 721\\
9. Not precise verb    & 2 & 111 & 27 & 26 & 33 & 32 & 231 \\
\bottomrule
\end{tabularx}
\end{table*}

Table~\ref{table:annotationRimayS1} shows the frequencies of Rimay patterns assigned by the annotators to requirements in $S$.
The first column shows the Rimay pattern. The second through seventh columns present the number of requirements assigned to each Rimay pattern for each batch in $S$. The last column displays the total number of occurrences of each of the 10 Rimay patterns in $S$.
Table~\ref{table:annotationRimayS1} suggests that, similar to smells, the SRSs used in our case study have significant diversity in their structural composition and patterns occur according to widely different frequencies.
For example, ``10. Multiple conditions and system response'' is the most frequently assigned Rimay pattern with 803 occurrences in set $S$. The least frequently assigned Rimay patterns (less than 20 times) are ``2. Scope, condition (precondition), and system response'' and ``8. Condition (Time) and system response''. In Table~\ref{table:annotationRimayS1}, we found that the pattern ``4. Scope, condition (Time) and system response'' is not observed in $S$. Recall from Section~\ref{subsec:rimay_patterns} that Rimay patterns represent valid sequences of Rimay concepts used to write requirements in Rimay. 
Even though the SRSs do not have requirements that can be rewritten by applying this pattern, we opted to keep the pattern in \TOOL to provide the complete list of Rimay patterns.

\begin{table*}[t]
\centering
\caption{Rimay patterns - Annotation results  for set $S$}
\label{table:annotationRimayS1}
\begin{tabularx}{\textwidth}{X >{\hsize=0.15\hsize}Y >{\hsize=0.15\hsize}Y >{\hsize=0.15\hsize}Y >{\hsize=0.15\hsize}Y >{\hsize=0.15\hsize}Y >{\hsize=0.15\hsize}Y >{\hsize=0.15\hsize}Y}
\toprule
   \textbf{Rimay Pattern} & \textbf{Batch 1} & \textbf{Batch 2} & \textbf{Batch 3} & \textbf{Batch 4} & \textbf{Batch 5} & \textbf{Batch 6} & \textbf{Total}  \\
\midrule
1. Scope and system response & 6 & 16 & 7 & 3 &  1 & 7 & 40\\
2. Scope, condition (precondition), and system response &  9 &  3 &  0 & 0 &  0 &  0 & 12\\
3. Scope, condition (trigger), and system response &  13 &  28 &  113 &  0 &  0 &  83 & 237\\
4. Scope, condition (time), and system response & 0 & 0 &  0 &  0 &  0 &  0 & 0\\
5. System response &  91 & 45 & 40 &  203 & 111 &  37 & 527 \\
6. Condition (precondition) and system response & 75 &  13 & 5 &  21 &  35 & 7 & 156\\
7. Condition (trigger) and system response & 116 & 158 &  73 &  202 &  105&  104 & 758\\
8. Condition (time) and system response & 0 &  2 &  4 & 0 &  0 & 0 & 6\\
9. Scope, multiple conditions, and system response & 5 &  29 &  86 &  0 &  4& 39 & 163  \\
10. Multiple conditions and system response & 63 & 310 &  79 & 25 &  120&  206 & 803\\
\bottomrule
\end{tabularx}
\end{table*}

%% file: evaluation_metrics.tex
\subsection{Metrics}
\label{subsec:qa_metrics}

Our analysis of the accuracy of \TOOL, when detecting smells and suggesting Rimay patterns, is based on the precision and recall metrics. We applied \TOOL to the subsets of requirements $S_D$ and $S_E$ to detect smells and suggest Rimay patterns. We compared these results against the ground truth $GT$ (Section~\ref{subsec:qa_case_study_design}) using  precision and recall. 
For each smell $s$ and Rimay pattern $p$, we first classified \TOOL's predictions for a requirement $r$ into the following categories:

\noindent\textbf{\textit{True positives (TP)}} are the correct predictions. In smell detection, a TP occurs when \TOOL detects the same smell $s$ as the ground truth $GT$ for requirement $r$. In pattern suggestion, a TP occurs when \TOOL assigns the same Rimay pattern $p$ as $GT$ to requirement $r$.

\noindent\textbf{\textit{False negatives (FN)}} are missed annotations. In smell detection, a missed annotation occurs when \TOOL fails to detect the smell $s$ that is annotated for requirement $r$ in the ground truth $GT$. In pattern suggestion, a missed annotation occurs when \TOOL fails to suggest the pattern $p$ that is annotated for $r$ in $GT$.

\noindent\textbf{\textit{False positives (FP)}} are misclassified annotations. In smell detection, an FP occurs when \TOOL incorrectly indicates the presence of the smell $s$ that is not annotated for requirement $r$ in the ground truth $GT$. In pattern suggestion, an FP occurs when \TOOL incorrectly suggests the Rimay pattern $p$ that is not annotated for $r$ in $GT$.

Next, for each detected smell (or assigned Rimay pattern), denoted by $i$, we calculated the precision~($P_i$) as $P_i = TP_i / (TP_i{+}FP_i)$ and the recall~($R_i$) as $R_i = TP_i/(TP_i{+}FN_i)$, where $TP_i$, $FP_i$, and $FN_i$ denote, respectively, the number of true positives, false positives, and false negatives in predicting $i$.

Furthermore, we calculated the overall precision as \texttt{Overall-P} $= {\sum_{i=1}^{l} TP_{i}}~/~{\sum_{i=1}^{l} (TP_{i}+FP_{i})}$, where $i, \ldots, l$ denote either the nine smells for smell detection or the 10 Rimay patterns for pattern suggestion. The overall recall was calculated as \texttt{Overall-R} $={\sum_{i=1}^{l} TP_{i}}~/~{ \sum_{i=1}^{l} (TP_{i}+FN_{i})}$.

%% file: evaluation_implementation.tex
\subsection{\TOOL Development}
\label{subsec:qa_implementation}

\begin{figure*}[t]
\centering
 \centerline{\includegraphics[width=0.95\linewidth]{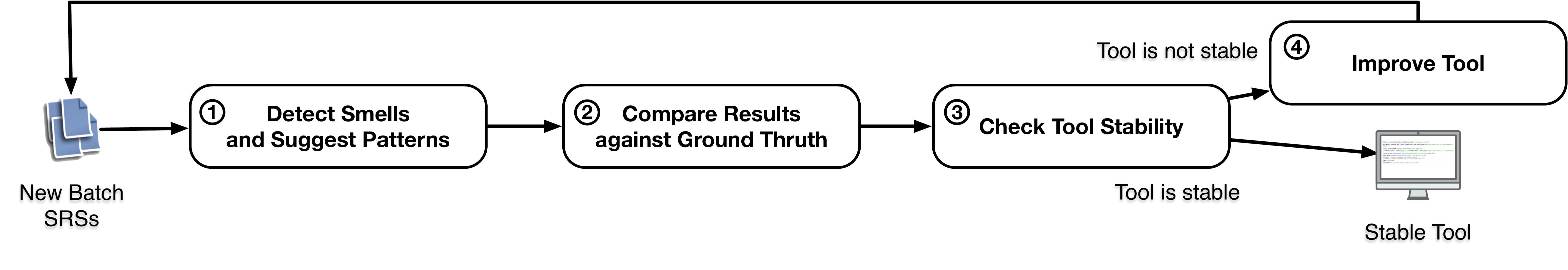}}
\caption{\TOOL development overview}
\label{fig:toolCreationOverview}
\end{figure*}

\begin{table}[t]
	\centering
	\caption{Batch distribution of the SRSs in $S$}
	\label{table:evalDatasetDistrib}
	\begin{tabularx}{\columnwidth}{l l Y Y}
		\toprule
		\textbf{Subset} & \textbf{Batch~\#} & \textbf{SRS ID} & \# \textbf{Requirements} \\
		\midrule
		\multirow{6}{*}{$S^D$} & \multirow{2}{*}{1} & SRS1 & 192 \\
		&  & SRS2 & 188 \\ 
		\arrayrulecolor{lightgray}\cline{2-4}\arrayrulecolor{black}
		&\multirow{2}{*}{2} & SRS3 & 118 \\
		& & SRS4 & 161 \\
		\arrayrulecolor{lightgray}\cline{2-4}\arrayrulecolor{black}
		&\multirow{2}{*}{3} & SRS5 & 451 \\
		& & SRS6 & 294 \\
		
		\arrayrulecolor{lightgray}\hline\arrayrulecolor{black}
		\multirow{7}{*}{$S^E$} & \multirow{2}{*}{4} & SRS7 & 367 \\
		& & SRS8 & 90 \\
		\arrayrulecolor{lightgray}\cline{2-4}\arrayrulecolor{black}
		&\multirow{3}{*}{5} & SRS9 & 167 \\
		& & SRS10 & 192 \\
		& & SRS11 & 19 \\
		
		\arrayrulecolor{lightgray}\cline{2-4} \arrayrulecolor{black}
		&\multirow{2}{*}{6} & SRS12 & 340 \\
		& & SRS13 & 146 \\            
		
		\bottomrule
	\end{tabularx}
\end{table}

\revision{We employed an iterative process to develop \TOOL, as depicted in Figure~\ref{fig:toolCreationOverview}.
For this iterative development,  we divided subset $S^D$ into three batches as detailed in Table~\ref{table:evalDatasetDistrib}.
Further, we also divided $S^E$ into three batches to prepare for the possibility that \TOOL would not become stable after using $S^D$, something difficult to predict at the development stage.
However, it turned out that \TOOL was stable after using $S^D$ and therefore no batches from $S^E$ were needed to reach a stable version of \TOOL.
}

As shown in Figure~\ref{fig:toolCreationOverview}, to build a stable version of \TOOL, we followed four steps: \textbf{(Step 1)} We applied \TOOL to a first batch of requirements listed in Table~\ref{table:evalDatasetDistrib} to detect smells and suggest Rimay patterns. \textbf{(Step 2)} We compared our results against the ground truth by computing precision and recall. \textbf{(Step 3)} We evaluated \TOOL's stability by applying the concept of saturation, which determines the point in the iterative development process of \TOOL where we have analyzed a sufficient number of SRSs to confidently identify smells and recommend Rimay patterns. In general, saturation refers to the point in a qualitative study when no new information emerges from the data being analyzed, i.e., when no new properties, dimensions, conditions, actions/interactions, or consequences are found in the data~\cite{GlaserSaturat}. In our context, during the development of \TOOL, we determined that the saturation point was reached when we observed stability from one batch of SRSs to the next in terms of overall precision and recall when identifying smells and recommended patterns. When such precision and recall were significantly worse than in the previous batch, we analyzed cases showing disagreements between \TOOL and the ground truth in order to account for new situations. \textbf{(Step 4)} We then improved \TOOL by correcting any disagreements with the ground truth. We repeated Steps 1 to 4 using each time a new batch as listed in Table~\ref{table:evalDatasetDistrib} until \TOOL became stable. For example, the results for batch 2 were initially much worse for smell detection than those of batch 1, whereas the results for batch 3 were right away comparable to those of batch 2 after improvement, thus suggesting no new situations needed to be investigated and accounted for. Below, we describe in detail how we improved \TOOL.

\begin{table*}[t]
\centering
\caption{Smell detection results for set $S^D$ (development) - precision and recall. The set $S^D$ contains batches 1, 2, and 3 listed in Table~\ref{table:evalDatasetDistrib}.}
\label{table:accuracySmellsTraining}
\begin{tabularx}{\textwidth}{X >{\hsize=0.2\hsize}Y >{\hsize=0.2\hsize}Y >{\hsize=0.2\hsize}Y >{\hsize=0.2\hsize}Y >{\hsize=0.2\hsize}Y >{\hsize=0.2\hsize}Y >{\hsize=0.2\hsize}Y >{\hsize=0.2\hsize}Y}
\toprule
\multirowcell{2}{\textbf{Smell}}&\multicolumn{2}{c}{\textbf{Batch 1}} & 
\multicolumn{2}{c}{\textbf{Batch 2}} & \multicolumn{2}{c}{\textbf{Batch 3}} & \multicolumn{2}{c}{\textbf{All ($S^D$)}} \\
\cmidrule(lr){2-3} \cmidrule(lr){4-5} \cmidrule(lr){6-7} \cmidrule(lr){8-9}
 &  P & R  & P & R & P & R & P & R\\
\midrule
1. Non-atomic                   & 1.00 & 1.00 & 0.92 & 0.92 & 0.94 & 0.94 & 0.94 & 0.94\\
2. Incomplete requirement       & 1.00 & 1.00 & 0.89 & 1.00 & 1.00 & 1.00 & 0.92 & 1.00\\
3. Incorrect order requirement  & 0.95 & 0.95 & 0.85 & 0.82 & 0.72 & 0.76 & 0.82 & 0.83\\
4. Coordination ambiguity       & N/A & N/A & 1.00 & 1.00 & N/A & N/A & 1.00 & 1.00 \\ 
5. Not a requirement            & 1.00 & 1.00 & N/A & N/A & N/A & N/A & 1.00 & 1.00\\ 
6. Incomplete condition         & 1.00 & 0.94 & 0.75 & 0.93 & 0.90 & 0.77 & 0.83 & 0.86 \\
7. Incomplete system response                & 1.00 & 1.00 & 1.00 & 0.89 & N/A & N/A & 1.00 & 0.91 \\ 
8. Passive voice                &  1.00 & 0.99 & 0.98 & 0.95 & 0.93 & 0.90 & 0.98 & 0.96\\
9. Not a precise verb          & 1.00 & 1.00 & 0.96 & 0.97 & 0.87 & 0.96 & 0.94 & 0.97 \\
        \midrule
        \midrule
        \textbf{Overall} & 0.99  & 0.99  & 0.92 & 0.94 & 0.90 & 0.88 & 0.93 & 0.93 \\
\bottomrule
\end{tabularx}
\end{table*}

Table~\ref{table:accuracySmellsTraining} shows precision ($P$) and recall ($R$) values obtained by \TOOL, after the iterative improvement mentioned above, for detecting smells in $S^D$. These results were calculated based on the ground truth $GT$. The first column of Table~\ref{table:accuracySmellsTraining} lists the smell names. 
The $P$ and $R$ values for each of the nine smells found in the first batch are shown in the second and third columns, while the fourth and fifth columns show $P$ and $R$ values for the smells found in the second batch. The sixth and seventh columns provide the $P$ and $R$ values for each of the smells found in the third batch of $S^D$. The overall $P$ and $R$ values for all three batches of $S^D$ are reported in the last two rows of Table~\ref{table:accuracySmellsTraining}.  

Following the iterative development process depicted earlier, we first developed \TOOL using the first batch of $S^D$ and compared our results with $GT$. In the first iteration, we obtained a precision ($P$) of 99\% and a recall ($R$) of 99\%. We then applied \TOOL to the second batch and obtained low $P$ and $R$ values (i.e., a $P$ of 69\% and an $R$ of 58\%),  which were considered unsatisfactory. To address this, we analyzed the false positives (FP) and false negatives (FN) (described in Section~\ref{subsec:qa_metrics}). Most of the FPs and FNs cases resulted from new scenarios that were not considered during the analysis of Batch 1. These new scenarios refer to requirements that require further investigation to improve \TOOL. We improved \TOOL to support these new scenarios and detect smells in the second batch, resulting in significant improvements with a precision of 92\% and a recall of 94\%. Further, we applied \TOOL to the third batch and obtained $P$ of 90\% and an $R$ of 88\%, which were deemed acceptable. These high precision and recall scores indicate that \TOOL detects most relevant smells in requirements with low probabilities of false positives and false negatives. At that point, we then stopped analyzing more SRSs and improving \TOOL as we reached saturation in terms of accuracy. 

\begin{table*}[t]
\centering
\caption{Rimay pattern suggestion results for set $S^D$ (development) - precision and recall. The set $S^D$ contains batches 1, 2, and 3 listed in Table~\ref{table:evalDatasetDistrib}.}
\label{table:resultsRimayTraining}
\begin{tabularx}{\textwidth}{X >{\hsize=0.13\hsize}Y >{\hsize=0.13\hsize}Y >{\hsize=0.13\hsize}Y >{\hsize=0.13\hsize}Y >{\hsize=0.13\hsize}Y >{\hsize=0.13\hsize}Y >{\hsize=0.13\hsize}Y >{\hsize=0.13\hsize}Y}
\toprule
\multirowcell{2}{\textbf{Rimay Pattern}} & \multicolumn{2}{c}{\textbf{Batch 1}} & \multicolumn{2}{c}{\textbf{Batch 2}}  & \multicolumn{2}{c}{\textbf{Batch 3}} & \multicolumn{2}{c}{\textbf{All ($S^D$)}} \\
\cmidrule(lr){2-3} \cmidrule(lr){4-5} \cmidrule(lr){6-7} \cmidrule(lr){8-9}
  & \textbf{P} & \textbf{R}  & \textbf{P} & \textbf{R} & \textbf{P} & \textbf{R} &\textbf{P} & \textbf{R}\\

\midrule
1. Scope and system response & 1.00 & 1.00 &  1.00 & 0.88 & 1.00 & 1.00 & 1.00 & 0.93 \\ 
2. Scope, condition (precondition), and system response &  0.80 & 0.89 & 1.00 & 1.00 & N/A & N/A & 0.79 & 0.92  \\ 
3. Scope, condition (trigger), and system response & 1.00 & 0.85 & 0.90 & 1.00 & 0.99 & 0.97 & 0.97 & 0.97 \\ 
4. Scope, condition (time) and system response & N/A & N/A & N/A & N/A & N/A & N/A & N/A & N/A \\
5. System response & 1.00 & 0.97 & 0.95 & 0.93 & 0.98 & 1.00 & 0.98 & 0.97 \\ 
6. Condition (precondition) and system response & 0.95 & 0.92 & 0.80 & 0.92 & 1.00 & 1.00 & 0.92 & 0.92 \\ 
7. Condition (trigger) and system response & 0.92 & 0.90 & 0.81 & 0.89 & 0.86 & 0.96 & 0.85 & 0.91\\ 
8. Condition (time) and system response & N/A & N/A & 1.00 & 1.00 & 1.00 & 0.75 & 1.00 & 0.83\\ 
9. Scope, multiple conditions, and system response & 0.83 & 1.00 & 1.00 & 0.93 & 0.89 & 0.84 & 0.91 & 0.87\\
10. Multiple conditions and system response & 0.95 & 0.89 & 0.96 & 0.87 & 0.83 & 0.78 & 0.94 & 0.86 \\
\midrule
\midrule
 \textbf{Overall} & 0.95 & 0.92 & 0.91 & 0.89 & 0.91 & 0.91 & 0.92 & 0.90\\ 
 \bottomrule
\end{tabularx}
\end{table*}

The precision ($P$) and recall ($R$) scores obtained by the stable version of \TOOL on $S^D$ when suggesting appropriate Rimay patterns to analysts, are presented in Table~\ref{table:resultsRimayTraining}. $P$ and $R$ values are reported for each Rimay pattern (column 1) suggested in the first, second, and third batches (columns 2-7), as well as the overall $P$ and $R$ scores for Batches 1, 2, and 3 in $S^D$ shown in the last row of the table.

Similarly to what we did for smell detection, we first developed \TOOL using the first batch of $S^D$ to recommend suitable Rimay patterns and compared our results with $GT$. The results were satisfactory. We obtained a precision ($P$) of 95\% and a recall ($R$) of 92\% in the first batch. Next, we applied \TOOL to the second and third batches, and the $P$ and $R$ values were around 90\%. We therefore reached saturation after the first batch, and no further enhancement to \TOOL was necessary. In terms of the overall accuracy, the suggestions for Rimay patterns yield an overall $P$ of 92\% and an $R$ of 90\% (Table~\ref{table:resultsRimayTraining}).

We note that Tables~\ref{table:accuracySmellsTraining} and \ref{table:resultsRimayTraining} show that in a few cases there is no  data to evaluate certain smells (such as ``5. Not a requirement'' from Table~\ref{table:accuracySmellsTraining}) and to suggest certain Rimay patterns (such as ``4. Scope, condition (time), and system response'' from Table~\ref{table:resultsRimayTraining}). This is indicated as N/A in the tables when the denominators of the precision and recall metrics are zero. We further discuss this issue in Section~\ref{sec:discussion}.

\textbf{\TOOL implementation and availability.} We implemented \TOOL using Python and Java. Specifically, we utilized spaCy~\cite{spacy}, Stanford CoreNLP~\cite{StanfordPostTag}, and AllenNLP~\cite{AllenNLP2017} to implement the preprocessing steps (Section~\ref{subsec:qa_preprocess}), including tokenization, post-tagging, and constituency parsing, respectively. Further, we employed the Stanford Tregex API for Java~\cite{LevyA06} to implement the Tregex patterns (Section~\ref{subsec:qa_separate_req}).
\TOOL is available online~\cite{Paska}.

%% file: evaluation_analys_data.tex
\subsection{Analysis and Results}
\label{sebsec:qa:analysis_results}

This section assesses the accuracy of \TOOL in detecting smells in NL requirements (RQ4) and suggesting Rimay patterns to analysts (RQ5).
We applied \TOOL to the evaluation set $S^E$ of requirements to detect smells and suggest Rimay patterns. 
Recall that $S^E$ was not used when developing \TOOL. 
We compared these results against the ground truth $GT$ (Section~\ref{subsec:qa_case_study_design}) using the precision and recall metrics (Section~\ref{subsec:qa_metrics}). 

\textbf{RQ4 results.} Table~\ref{table:accuracySmells} shows precision ($P$) and recall ($R$) values when detecting smells (RQ4). 
The first column of Table~\ref{table:accuracySmells} shows the smell name. The second and third columns show the $P$ and $R$ values for each of the nine smells found in the fourth batch. The fourth and fifth columns show the $P$ and $R$ values for each of the nine smells found in the fifth batch. The sixth and seventh columns show the $P$ and $R$ values for each of the nine smells found in the sixth batch of $S^E$. The last row of Table~\ref{table:accuracySmells} shows the overall $P$ and $R$ values for batches 4, 5, and 6 of $S^E$.

\begin{table*}[t]
\centering
\caption{Smell detection results for set $S^E$ (evaluation) - precision and recall. The evaluation set $S^E$ contains batches 4, 5, and 6 listed in Table~\ref{table:evalDatasetDistrib}.}
\label{table:accuracySmells}
\begin{tabularx}{\textwidth}{X >{\hsize=0.2\hsize}Y >{\hsize=0.2\hsize}Y >{\hsize=0.2\hsize}Y >{\hsize=0.2\hsize}Y >{\hsize=0.2\hsize}Y >{\hsize=0.2\hsize}Y >{\hsize=0.2\hsize}Y >{\hsize=0.2\hsize}Y}
\toprule
\multirowcell{2}{\textbf{Smell}}&\multicolumn{2}{c}{\textbf{Batch 4}} & 
\multicolumn{2}{c}{\textbf{Batch 5}} & \multicolumn{2}{c}{\textbf{Batch 6}} & \multicolumn{2}{c}{\textbf{All ($S^E$)}} \\
\cmidrule(lr){2-3} \cmidrule(lr){4-5} \cmidrule(lr){6-7} \cmidrule(lr){8-9}
 &  P & R  & P & R & P & R & P & R\\
\midrule
1. Non-atomic                   & 0.71  & 0.83  & 1.00 & 0.83   & 0.88 & 0.92 & 0.89 & 0.90 \\
2. Incomplete requirement       & 1.00  & 1.00   & 1.00   & 1.00   & 1.00 & 1.00 & 1.00 & 1.00 \\
3. Incorrect order requirement  & 0.90  & 1.00  & 0.88  & 1.00  & 0.74 & 1.00 & 0.81 & 1.00 \\
4. Coordination ambiguity       &  1.00  & 1.00   & 0.80  & 0.80  & 0.80 & 1.00 & 0.82 & 0.90 \\
5. Not a requirement            & 1.00  & 1.00  & 0.50   & 1.00   & 1.00 & 1.00 & 0.75 & 1.00\\
6. Incomplete condition         &  0.90 & 1.00 & 0.75 & 0.67    & 0.85 & 0.87 & 0.82 & 0.81 \\
7. Incomplete system response                &  1.00 & 1.00 & 0.33 & 1.00    & 0.67 & 1.00 & 0.62 & 1.00\\
8. Passive voice                & 0.80 & 0.85 & 0.91 & 0.91 & 0.99 & 0.91 & 0.94 & 0.90 \\
9. Not a precise verb          &  0.96 & 0.88 & 0.97 & 0.85 & 1.00 & 1.00 & 0.98 & 0.91\\
        \midrule
        \midrule
        \textbf{Overall} & 0.87 & 0.90 &  0.88 & 0.84 & 0.91 & 0.91 & 0.89 & 0.89\\
\bottomrule
\end{tabularx}
\end{table*}

As shown in Table~\ref{table:accuracySmells}, we achieved a precision ($P$) of 87\% and a recall ($R$) of 90\% for the fourth batch, a $P$ of 88\% and an $R$ of 84\% for the fifth batch, and a $P$ of 91\% and an $R$ of 91\% for the sixth batch. Overall, the detection of smells yields $P$ = $R$ = 89\%. 

\begin{tcolorbox}[enhanced jigsaw,left=2pt,right=2pt,top=0pt,bottom=0pt,opacityback=0]
\emph{The answer to RQ4 is that}, on a large number of real requirements from the financial domain, \TOOL shows a high degree of accuracy in detecting smells in NL requirements, with an overall precision and recall of 89\%. Therefore, \TOOL detects most smells in NL requirements with a small number of false positives and false negatives.
\end{tcolorbox}

From the data in Table~\ref{table:accuracySmells}, we observe that there are low precision ($P$) scores in the Batch 5 (i.e., ``5. Not a requirement'' with $P$ = 50\% and ``7. Incomplete system response'' with $P$ = 33\%) and Batch 6 (i.e., ``7. Incomplete system response'' with $P$ = 67\%). Furthermore, we observe a low recall ($R$) value in Batch 5 (i.e., ``6. Incomplete condition'' with an $R$ of 67\%). These are mainly caused by new scenarios that were not observed in the SRSs ($S^D$) used when developing \TOOL and limitations of the NLP tools employed by \TOOL.
We further discuss the reasons for such low precision and recall cases in Section~\ref{sec:discussion}.


\begin{table*}[t]
\centering
\caption{Rimay pattern suggestion results for set $S^E$ (evaluation) - precision and recall. The evaluation set $S^E$ contains batches 4, 5, and 6 listed in Table~\ref{table:evalDatasetDistrib}.}
\label{table:resultsRimay}
\begin{tabularx}{\textwidth}{X >{\hsize=0.13\hsize}Y >{\hsize=0.13\hsize}Y >{\hsize=0.13\hsize}Y >{\hsize=0.13\hsize}Y >{\hsize=0.13\hsize}Y >{\hsize=0.13\hsize}Y >{\hsize=0.13\hsize}Y >{\hsize=0.13\hsize}Y}
\toprule
\multirowcell{2}{\textbf{Rimay Pattern}} & \multicolumn{2}{c}{\textbf{Batch 4}} & \multicolumn{2}{c}{\textbf{Batch 5}}  & \multicolumn{2}{c}{\textbf{Batch 6}} & \multicolumn{2}{c}{\textbf{All ($S^E$)}} \\
\cmidrule(lr){2-3} \cmidrule(lr){4-5} \cmidrule(lr){6-7} \cmidrule(lr){8-9}
  & \textbf{P} & \textbf{R}  & \textbf{P} & \textbf{R} & \textbf{P} & \textbf{R} &\textbf{P} & \textbf{R}\\

\midrule
1. Scope and system response & 1.00 & 1.00  & 1.00 & 1.00 & 1.00 & 1.00 & 1.00 & 1.00 \\ 
2. Scope, condition (precondition), and system response & N/A & N/A & N/A   & N/A  & N/A & N/A & N/A & N/A  \\ 
3. Scope, condition (trigger), and system response & N/A & N/A & N/A & N/A & 0.98 & 1.00 & 0.98 & 1.00\\ 
4. Scope, condition (time) and system response & N/A & N/A & N/A & N/A & N/A & N/A & N/A & N/A\\
5. System response &  1.00 & 0.99 & 0.98 & 1.00 & 0.97 & 1.00 & 0.99 & 0.99 \\ 
6. Condition (precondition) and system response & 1.00 & 0.95  & 0.89 & 0.94 & 0.75 & 0.86 & 0.91 & 0.94\\ 
7. Condition (trigger) and system response & 0.98 & 0.93 & 0.86 & 0.83 & 0.99 & 0.87 & 0.95 & 0.89\\ 
8. Condition (time) and system response & N/A & N/A & N/A & N/A & N/A & N/A & N/A & N/A\\ 
9. Scope, multiple conditions, and system response & N/A & N/A & 0.80 & 1.00 & 0.94 & 0.79 & 0.92 & 0.81 \\
10. Multiple conditions and system response & 0.76 & 1.00 & 0.96 & 0.85 & 0.94 & 0.99 & 0.93 & 0.94\\
\midrule
\midrule
 \textbf{Overall} & 0.97 & 0.96 & 0.93 & 0.90 & 0.96 & 0.95 & 0.96 & 0.94\\ 
 \bottomrule
\end{tabularx}
\end{table*}

\textbf{RQ5 results.} Table~\ref{table:resultsRimay} shows the precision ($P$) and recall ($R$) scores obtained by \TOOL  when suggesting suitable Rimay patterns to analysts (RQ5). Once again, the results were calculated by applying \TOOL to the evaluation set $S^E$ and comparing the results to the ground truth $GT$ by using the precision and recall metrics (Section~\ref{subsec:qa_metrics}). 
Table~\ref{table:resultsRimay} shows, for all Rimay patterns (column 1), the $P$ and $R$ values for the patterns suggested in Batch 4 (columns 2-3), Batch 5 (columns 4-5), Batch 6 (columns 6-7), and all batches (columns 8-9). 
The last row of Table~\ref{table:resultsRimay} shows the overall scores of $P$ and $R$ for batches 4, 5, and 6 in $S^E$.
We obtained $P$ = 97\% and $R$ = 96\% in Batch 4,  $P$ = 93\% and $R$ = 90\% in Batch 5, and $P$ = 96\% and $R$ = 95\% in Batch 6. In terms of overall accuracy, the suggestions for Rimay patterns yield an overall $P$ = 96\% and $R$ = 94\%~(Table~\ref{table:resultsRimay}).

We note that Table~\ref{table:resultsRimay} shows the results obtained by applying \TOOL to all 1321 requirements in $S^E$, including requirements with and without smells. \TOOL is indeed applicable for consistently rewriting requirements in Rimay, with or without smells. If we consider only the requirements with smells in $S^E$, we have a subset of 1165 such requirements for which we obtained $P$ = 90\% and $R$ = 87\%. 

\begin{tcolorbox}[enhanced jigsaw,left=2pt,right=2pt,top=0pt,bottom=0pt,opacityback=0]
\emph{The answer to RQ5 is that} \TOOL is accurate in suggesting Rimay patterns, achieving an overall precision of 96\% and recall of 94\%. Hence, most of the time, \TOOL provides analysts with appropriate Rimay patterns to fix smells in NL requirements, with relatively few false positives and false negatives.
\end{tcolorbox}

As shown in Table~\ref{table:resultsRimay}, there are low precision ($P$) values in Batch 4 (i.e., ``10. Multiple conditions and system response'' with a $P$ of 76\%) and the Batch 6 (i.e., ``6. Condition (precondition) and system response'' with a $P$ of 75\%). These low values are mainly caused by new scenarios that were not observed in the SRSs $S^D$ used when developing \TOOL.
In Section~\ref{sec:discussion}, we further examine the factors that lead to low $P$ values. 
In addition, we found that some Rimay patterns are not applicable to any of the requirements in the SRSs $S^E$, as indicated in Table~\ref{table:resultsRimay} by N/A, e.g., ``2. Scope, condition (precondition) and system response''. Section~\ref{sec:discussion} further discusses such situations.

\textbf{Remark.}
In our evaluation results, we observed that \TOOL obtained high overall precision and recall scores: respectively, 89\% and 89\% for smell detection, and 96\% and 94\% for pattern suggestion.
Even though we would ideally like to achieve 100\% accuracy, our results are promising and demonstrate the potential of \TOOL to support requirements quality assurance.

%% file: discussion.tex
\section{Discussion}
\label{sec:discussion}

\subsection{Approach Performance.} The results of RQ4 presented in Section~\ref{sec:qa_evaluation} show that \TOOL is accurate in terms of detecting smells in NL requirements (P = 89\% and R = 89\%). However, we also observed that it achieved low precision scores for the detection of particular smells, ``5. Not a requirement'' (50\% in Batch~5), ``7. Incomplete system response'' (33\%  in the Batch~5 and 67\% in the Batch~6), and a low recall score for detecting the smell ``6. Incomplete condition'' (67\% in the Batch~5). To determine the root causes of such low precision and recall values, we analyzed the false positives and false negatives for each of the smells mentioned above.
Note that, for the smell ``5. Not a requirement'' in Batch~5, our results show 1 true positive, 1242 true negatives, 1 false positive, and no false negatives.
For the smell ``7. Incomplete system response'' in Batch~5,  we have 1 true positive, 1241 true negatives, 2 false positives, and no false negatives.
Regarding the smell ``7. Incomplete system response'' in Batch~6, the results have 2 true positives, 2222 true negatives, 1 false positive, and no false negatives.
For the smell ``6. Incomplete condition'' in Batch~5, we obtained 33 true positives, 1184 true negatives, 11 false positives, and 16 false negatives.

\begin{table*}[t]
\centering

\caption{Examples of (anonymized) requirements that lead to \TOOL failing to accurately detect smells and suggest Rimay patterns.}
\label{tab:error_Requirements}
\begin{tabularx}{\textwidth}{>{\hsize=0.05\hsize}X X}
\toprule
\hfil \textbf{ID} &  \hfil \textbf{Requirement Description} \\
\midrule
\hfil \texttt{R1} & \exCNL{The System-A must route the outbound messages to System-B instead of System-C.}\\
\arrayrulecolor{lightgray}\hline\arrayrulecolor{black}
\hfil \texttt{R2} & \exCNL{Upon receipt of a valid C01 cancellation from System-A Participant, then the System-B must route the cancellation to the same destination.}\\

\arrayrulecolor{lightgray}\hline\arrayrulecolor{black}
\hfil \texttt{R3} & \exCNL{if the System-A Order Issuer Ordering data = Value-A}\\
\arrayrulecolor{lightgray}\hline\arrayrulecolor{black}
\hfil \texttt{R4} & \exCNL{When the user clicks on the Edit icon of Screen-1, System-A must set in updatable mode the following fields:} \textbullet~\exCNL{Include portfolio in the S-Order,} \textbullet~\exCNL{Alert to operations (e.g., when Order is rejected), ...}\\


\arrayrulecolor{lightgray}\hline\arrayrulecolor{black}
\hfil \texttt{R5} & \exCNL{If the Participant Status = Delete, then System-A must populate the field Status with the value inactive.}\\
\bottomrule
\end{tabularx}
\end{table*}

\textbf{\textit{5. Not a requirement.}} Recall from Section~\ref{subsec:smells} that this smell occurs when there is a statement that does not contain any requirement segment, i.e., scope, condition, and system response. 
We observed that Batch~5 has few requirements related to the smell ``5. Not a requirement''. For the smell, Batch~5 has 1 true positive, 1242 true negatives, 1 false positive, and no false negatives, thus giving a low precision of 50\%. The false positive (i.e., incorrectly detected smells) was related to inaccurate POS tags assigned to the verb of the system response. Table~\ref{tab:error_Requirements} shows requirement R1. This requirement has only a system response. The verb of the system response of R1 (i.e., route) was incorrectly identified as a noun. Since no verb was found in the system response of the requirement, \TOOL detected the smell ``5. Not a requirement''.

    
\textit{\textbf{6. Incomplete Condition.}} Recall from Section~\ref{subsec:smells} that this smell occurs when the condition of the requirement misses the verb or the actor. We observed that the cause of false negatives (i.e., \TOOL failed to detect smells) was related to scenarios not observed during the development of \TOOL. For the smell ``6. Incomplete condition'', Batch~5 has 33 true positives , 1184 true negatives, 11 false positives, and 16 false negatives, resulting in a low recall of 67\%. 
All false negatives were related to new scenarios that were not observed during the development of \TOOL. Table~\ref{tab:error_Requirements} shows an example (R3) of a false negative. The condition R3 misses a verb and instead has the symbol ``='', which denotes ``equals to''. Furthermore, R3 is made up of a compound noun, ``System-A Order Issuer Ordering data''. According to the POS Tagger, the word ``Ordering'' of the compound noun is recognized as a verb, indicating that the condition is complete. Therefore, \TOOL did not trigger any smells. However, R3 actually misses a verb.


\textit{\textbf{7. Incomplete system response.}} Recall from Section~\ref{subsec:smells} that this smell occurs when the system response misses the actor, the modal verb or the verb. We noted that the cause of false positives (i.e., \TOOL incorrectly detected smells) was the assignment of incorrect POS tags to the verbs in the system response. Such absence of verb in the system response triggers the smell ``7. Incomplete system response”. 
Regarding the smell, Batch~5 has 1 true positive, 1241 true negatives, 2 false positives, and no false negatives, resulting in a low precision of 33\%.  Furthermore, for the same smell, Batch~6 has 2 true positives, 2222 true negatives, 1 false positive, and no false negatives, resulting in a low precision of 67\%.
We found that all false positives were related to incorrect POS tags assigned to the verb in the system response. Table~\ref{tab:error_Requirements} shows requirement R2. The verb of the system response of R2 (i.e., route) was incorrectly identified as a noun. Since no verb was found in the system response of the requirement, \TOOL incorrectly triggered the smell ``7. Incomplete system response''. 

Similarly, the results of RQ5 show that \TOOL accurately suggests requirement patterns in most cases ($P$ = 96\% and $R$ = 94\%). However, we noted that low precision scores were obtained by \TOOL when suggesting requirement patterns ``10. Multiple conditions and system response” (76\% in the Batch~4) and ``6. Condition (precondition) and system response” (75\% in the Batch~6). To determine the root causes of such low precision, we analyzed cases in which \TOOL yielded false positives (i.e., incorrectly suggested patterns).
Note that, for the pattern ``10. Multiple conditions and system response’’, Batch~4 has 25 true positives, 421 true negatives, 8 false positives, and no false negatives. Regarding the pattern ``6. Condition (precondition) and system response’’, Batch~6 has 6 true positives, 474 true negatives, 2 false positives, and 1 false negative.

\textbf{\textit{Pattern: 10. Multiple conditions and system response.}} This pattern is suggested when a requirement has the following segments: two or more conditions and a system response. 
For the pattern ``10. Multiple conditions and system response’’, Batch~4 has 25 true positives, 421 true negatives, 8 false positives, and no false negatives, thus obtaining a low precision of 76\%. 
The majority of false positives (i.e., six out of eight) were due to new scenarios that were not observed during the development of \TOOL. The remaining two false positives were related to human errors in the annotation. 
Requirement R4 in Table~\ref{tab:error_Requirements} shows an example that causes \TOOL to incorrectly suggest a Rimay pattern. R4 has a condition with a system response containing bullet points. The second bullet point contains the condition ``when Order is rejected'', which only applies to the information in the bullet point. However, \TOOL incorrectly identified it as a condition that applies to the entire requirement; therefore, \TOOL suggested the Rimay pattern ``10. Multiple conditions and system response.'' Nevertheless, in reality, R4 has only one condition. 


\textbf{\textit{Pattern: 6. Condition (precondition) and system response}}. This pattern is suggested when a requirement has the following segments: a condition of type precondition and a system response. 
For the pattern ``6. Condition (precondition) and system response’’, Batch~6 has 6 true positives, 474 true negatives, 2 false positives, and 1 false negative, resulting in a low precision of 75\%.
False positives were related to new scenarios that were not observed during the development of \TOOL. 
Requirement R5, in 
Table~\ref{tab:error_Requirements}, is an example of a false positive (R5). The condition of R5 lacks a verb; instead, R5 has the operator ``='' which denotes ``equals to''. Furthermore, R5 is made up of a compound noun, ``Participant Status''. The word ``Status'' of the compound noun is identified as a verb by the POS Tagger which suggests that the condition is a condition of type trigger (Section~\ref{subsec:qa_rimay_patterns}). Therefore, \TOOL suggested the Rimay pattern ``7. Condition (trigger) and system response''. However, the condition of R5 is a condition of type precondition because of the symbol ``='' which denotes ``equals to''.


In summary, we identified two main reasons for low precision and recall in the cases mentioned above. The first is POS Tagger limitations: POS Tagger incorrectly assigns POS tags to words, which causes \TOOL to incorrectly identify the smells and syntax of the requirement. \TOOL does not have control over the accuracy of the POS Tagger, since it is a third-party component. Second, we found several new scenarios that were not previously observed. These scenarios include different structures of the requirement segments and the presence of additional information in requirements. We could enhance \TOOL to support such new scenarios. However, some of them are examples of bad practices in specifying requirements. For example, in requirement R4 of Table~\ref{tab:error_Requirements}, the analyst has inserted a condition as an additional information in the system response.
In requirement R5, the symbol ``='' is used. However, in general, using such symbols may result in ambiguous interpretations among stakeholders. Hence, it is recommended to avoid using such symbols when writing requirements to prevent confusion.

\subsection{Lack of Testing Data}
We observed  during the development of \TOOL (Table~\ref{table:accuracySmellsTraining}) the lack of testing data to evaluate some smells for certain batches (i.e., ``4. Coordination ambiguity'' in batches 1 and 3, ``5. Not a requirement'' in batches 2 and 3, and ``7. Incomplete system response'' in Batch~3). We were unable to test \TOOL in the above cases because $S^D$ lacks requirements that contain these smells. However, the aforementioned cases did not occur in all batches of $S^D$. The smells detected by \TOOL were all tested in at least  one batch during the development of \TOOL. 
We also noted that test data were missing to evaluate the suggestion of some Rimay patterns (Table~\ref{table:resultsRimayTraining}): ``2. Scope, condition (precondition), and system response'' in Batch~3, ``4. Scope, condition (time) and system response'' in batches 1 to 3, and ``8. Condition (time) and system response'' in Batch~1.
However, the Rimay patterns ``2. Scope, condition (precondition), and system response'' and ``8. Condition (time) and system response'' were tested in other batches in $S^D$ during the development of \TOOL.
Regarding the pattern ``4. Scope, condition (time) and system response'', although the SRSs do not have requirements that can be rewritten by applying the pattern, we opted to keep the pattern in \TOOL to support the complete list of Rimay patterns.
Recall from Section~\ref{subsec:rimay_patterns} that Rimay patterns represent valid sequences of Rimay concepts used to write requirements in Rimay.

Similarly, the results of RQ5 (Table~\ref{table:resultsRimay}) show that \TOOL was not tested when suggesting the Rimay patterns ``2. Scope, condition (precondition), and system response” in batches 4 to 6, ``3. Scope, condition (trigger), and system response'' in batches 4 and 5, ``4. Scope, condition (Time), and system response” in batches 4 to 6, ``8. Condition (time) and system response'' in batches 4 to 6, and ``9. Scope, multiple conditions, and system response'' in Batch~4.
\TOOL could not be tested in the above cases because $S^E$ did not include requirements with the syntax necessary to suggest the corresponding Rimay patterns. As described in Section~\ref{subsec:qa_data_collection}, the 13 SRSs used to evaluate \TOOL were collected from our industrial partner, who deemed them representative of recent SRSs and over which we had no control. 

%% file: threats.tex
\section{Threats to Validity}
\label{sec:qa_treats}

\textbf{Internal validity} is of concern when examining causal relations~\cite{Runeson:2012}. Our results depend heavily on the quality of annotations, which are susceptible to annotation biases. To minimize any potential biases, we hired three external annotators who did not have access to \TOOL in our experiments. To ensure the high quality of their annotations, we provided training sessions and monitored the annotation agreement between annotators using Cohen's Kappa metric.
To minimize any biases introduced by our monitoring activities, we limited our inspection of their annotations to the requirements for which the annotators indicated having difficulty with. In addition, at the final stage of the annotation process, we randomly selected 10\% of the annotated requirements for inspection.

In our experiments, two external annotators were employed to annotate Rimay patterns. For smells, however, another external annotator and the first author of this article were responsible for annotating the SRSs. To mitigate any experimenter bias introduced by the first author, we included the author's annotations only in the development set (i.e., $S^D$ described in Section~\ref{subsec:qa_data_collection}) that was used to develop \TOOL. Hence, our results obtained from the evaluation set (i.e., $S^E$ described in Section~\ref{subsec:qa_data_collection}) were not impacted by the first author's annotations.

\revision{
Another threat to internal validity concerns potential biases introduced by specific researchers.
Recall from Section~\ref{subsec:smells} that the first author of this article defined a catalog of nine smells that \TOOL detects.
To mitigate this threat, we validated the nine smells with our industrial partner to ensure their relevance to the errors commonly observed in NL requirements writing.
Further, the other authors of this article closely monitored the smell identification process.
}

\revision{
Recall from Section~\ref{subsec:qa_implementation} that we applied the concept of saturation to develop a stable version of \TOOL.
To make this process rigorous and objective, we measured precision and recall after each batch, using a separate test dataset.
The saturation point was considered to be reached when we started to observe consistent overall precision and recall from one batch of SRSs to the next.
}

\revision{
\TOOL employs NLP techniques such as tokenization, POS tagging, and constituency parsing, over which it does not have direct control regarding their accuracy.
Recognizing the importance of such accuracy and accounting for the latest advances of these techniques, we implemented \TOOL using well-maintained, state-of-the-art NLP libraries: spaCy~\cite{spacy}, Stanford CoreNLP~\cite{StanfordPostTag}, and AllenNLP~\cite{AllenNLP2017}.
Further, these libraries have also been widely applied in various domains~\cite{FemmerFWE17,OsamaZAGI20,EzziniA0S22}.
In the end, our choice of techniques and libraries led to an implementation showcasing high precision and recall, indicating that \TOOL is a promising solution.
Nevertheless, we note that supplemental preprocessing of NL requirements (e.g., grammar corrections), which can be executed independently of \TOOL, has the potential to enhance the accuracy of these NLP techniques.
Such enhancements could thus further optimize the performance of \TOOL.
}

\textbf{External validity} concerns the degree to which our results can be generalized to other contexts~\cite{Runeson:2012}. 
In our experiments, we evaluated \TOOL using industrial SRSs that contain NL requirements from 13 systems (SRSs) in the financial domain. Specifically, out of the 13 SRSs, six ($S^D$) were used for developing \TOOL, and the remaining seven ($S^E$) were used in our evaluation to answer RQ4 and RQ5. These requirements are however representative of a broader class of information systems, such as those used by our industrial partner for data management, security compliance, and communication. Furthermore, the requirements were written by different analysts with different backgrounds, which increases the diversity of the SRSs.
\revision{We note that, of the 13 SRSs, six were previously used to develop the Rimay language~\cite{VeizagaATSB21}.
From these six SRSs, two were allocated to the development set $S^D$ and four to the evaluation set $S^E$.
For this study, we obtained seven new SRSs from our industrial partner.
The overlap in SRSs used to define Rimay and to develop and assess \TOOL is not a threat to validity since, in this work, we assume Rimay is already valid and complete, as previously studied, and here we only assess our ability to detect smells and recommend patterns, clearly dividing SRSs into development and test sets.}
Though our results should be generalizable to information systems in other domains, future investigations are nevertheless necessary to determine how \TOOL fares outside finance.
\revision{In the future, despite the large number of requirements we used in our study, when working on other SRSs, we might uncover patterns we have not identified yet. This would require that we augment or modify the smells related to our catalog of patterns.}

Recall from Section~\ref{sec:qa_approach} that, to identify the beginning of each segment and imprecise verbs in a requirement, \TOOL uses keyword-based analysis techniques.
These techniques rely on glossaries that are created based on the keywords defined in Rimay and our inspection of the representative SRSs used in our study. Furthermore, our industrial partner validated the glossaries. 
However, further research is needed in order for \TOOL to rely on more complete glossaries. Due to the simplicity of the techniques, one can easily expand our glossaries by leveraging those defined in existing work (e.g., Smella~\cite{FemmerFWE17} for detecting requirements smells).

\revision{
Regarding the use of Tregex, which defines regular expression-like patterns on the syntax tree of a requirement, incorrectly defined patterns may miss the syntax sub-trees they aim to identify.
In \TOOL, these inaccuracies can result in false positives and false negatives when detecting smells and suggesting recommendations.
To ensure that we correctly defined the Tregex patterns, we used the development set $S^D$, which contains diverse requirements written by various analysts for six different systems, as a basis for defining and verifying the patterns.
In addition, we inspected the false positives and false negatives from our experiment results on $S^E$, which show that \TOOL detected smells with a precision of 89\% and a recall of 89\%, while suggesting recommendations with a precision of 96\% and a recall of 94\%.
This careful inspection indicates that these false positives and false negatives come from incorrect POS tags, the inclusion of symbols (e.g., ``='') in a requirement, and annotation errors, rather than from any potential mistakes we might have introduced in the Tregex patterns.
Nevertheless, further studies that analyze actual requirements from different domains are needed to assess the completeness of the Tregex patterns.
}

%% file: related.tex
\section{Related Work}
\label{sec:qa_related}

In this section, we compare our work with existing studies related to improving the quality of NL requirements. In particular, we will discuss research strands in the areas of assisting analysts in (1)~writing requirements based on templates, (2)~assuring the quality of requirements by analyzing incompleteness and ambiguity in requirements, and (3)~identifying smells in requirements.

\textbf{Requirements templates.} Requirements templates (e.g., EARS~\cite{mavin2009easy} and Rupp~\cite{Pohl11}) have been widely used in many studies and in practice~\cite{mavin2010big,AroraSBZ15,MavinWGU16,SleimiCSBD20,VeizagaATSB21}. These templates provide a structured approach to writing requirements, which can help reduce ambiguity, increase clarity, and ensure consistency across requirements. However, since requirements templates only provide coarse-grained concepts and constructs (e.g., the EARS condition template \texttt{WHEN <event>} does not specifically specify the content of the  event that initiates the requirement, allowing analysts to introduce free text), only a limited number of automated analysis techniques that rely on such templates have been introduced~\cite{AroraSBZ15}. \citet{AroraSBZ15} presented an automated approach for checking conformance to requirements templates (e.g., EARS and Rupp). The approach relies on an NLP technique, known as text chunking. In contrast, we rely on Rimay, a state-of-the-art CNL for specifying requirements. Rimay provides structures with fine-grained concepts and constructs that enabled us to develop an automated tool (\TOOL) for effectively detecting requirements smells and providing Rimay patterns as recommendations for removing smells in requirements. 

\textbf{Quality assurance.} Among the many strands of quality assurance research~\cite{Denger2005} in requirements engineering, the most pertinent ones introduce automated methods for detecting quality problems in requirements. In particular, we discuss prior work that aims at detecting problems of incompleteness and ambiguity in requirements.

\emph{Incompleteness analysis.} Completeness of requirements is often viewed from two perspectives: external and internal~\cite{Zowghi2003OnTI}. External completeness ensures that all necessary functionalities of a system are specified in the requirements.
Many of the existing studies on the completeness of requirements belong to this research strand~\cite{CostelloL95,KaiyaS05,FerraridSG14,DalpiazSL18,AroraSB19}. For example, \citet{DalpiazSL18} combined NLP and information visualization techniques to identify missing requirements. Their approach relies on the notion of stakeholders' viewpoints, which helps analysts identify cases in which one viewpoint mention concepts that are not present in other viewpoints.
\citet{AroraSB19} empirically evaluated the usefulness of domain models in detecting incompleteness of requirements. They conducted experiments by seeding some omissions to the requirements and checked whether the domain model can be used to detect the omissions.
In contrast to these prior studies, our work is related to the research strand on internal completeness of a requirement, which is concerned with ensuring that the requirement is self-contained. This means that the requirement contains all the information contents required to express its function completely.
Recent smell detection work~\cite{FemmerFWE17,FerrariGRTBFG18} include techniques to detect missing contents in a requirement, such as measurement units and references. In contrast, \TOOL is able to identify different information contents that are missed in a requirement: i.e., (1)~missing actors or verbs in conditions, (2)~missing actors, modal verbs, or verbs in system requirements, and (3)~missing system responses in requirements, as it relies on Rimay's concepts and constructs.
We further discuss these smell detection approaches below in the smell detection paragraphs.

\emph{Ambiguity analysis.} Ambiguity is a persistent issue in NL requirements. Hence, such ambiguity has been extensively studied in the literature~\cite{KiyavitskayaZMB08,YangRGWN11,OsamaZAGI20,EzziniA0SB21,EzziniA0S22}. For example, recently, \citet{EzziniA0S22} proposed six alternative solutions for automating the handling of anaphoric ambiguity in requirements. These solutions incorporate both traditional and state-of-the-art NLP and ML techniques, such as SpanBERT~\cite{JoshiCLWZL20}. \citet{OsamaZAGI20} introduced a technique for detecting syntatic ambiguity in a requirement using scored interpretations of the requirement, which provide users with most likely interpretations.
More precisely, the technique relies on an NLP algorithm that generates parsing trees with confidence scores to provide scored interpretations of a requirement.
However, the research strands in this area differ from our work, which focuses on identifying and addressing requirements smells. 

While incompleteness and ambiguity are types of requirements smells, our study aims at detecting a broader range of smells that indicate quality problems in requirements and providing recommendations to solve them.
When a requirement has multiple smells, analysts need to account for them together to obtain a complete picture of the overall quality of the requirement, a necessary condition to fix it properly. \TOOL helps analysts in this process by automatically detecting smells and suggesting patterns to fix them.

\textbf{Smell detection.} The research strands most related to our work are those that aim at detecting smells in NL text, such as requirements, feature requests, and use-case descriptions. Below, we discuss recent studies in this research area: two studies on NL requirements~\cite{FemmerFWE17,FerrariGRTBFG18}, and two others on feature requests~\cite{MuSZZZ20} and use-case descriptions~\cite{SekiHS19}. All these studies shared the objectives of defining catalogs of smells in NL descriptions and presenting automated smell detection techniques.

\citet{FemmerFWE17} introduced a tool, named Smella, that detects nine smells in NL requirements: subjective language, ambiguous adverbs and adjectives, loopholes, non-verifiable terms, superlatives, comparatives, negative words, vague pronouns, and incomplete references. Smella relies on POS tagging, dictionaries, and lemmatization. \citet{FemmerFWE17} evaluated Smella on 336 requirements, 53 use cases, and 1082 user stories collected from three companies and several university students. 
Their evaluation results showed that Smella achieved, on average, a precision of 59\% and a recall of 82\%.

\citet{FerrariGRTBFG18} presented an approach that detects the following requirement smells: anaphoric ambiguity, coordination ambiguity, vague terms, modal adverbs, passive voice, excessive length, missing condition, missing unit of measurement, missing reference, and undefined term. To detect these smells, they defined a set of smell-detection patterns, i.e., sequences of tokens to be matched within a requirement, relying on NLP techniques such as tokenization, POS tagging, and shallow parsing. They applied the patterns to 1866 requirements obtained from a railway company and obtained a precision of 83\% and a recall of 85\%.

\citet{MuSZZZ20} introduced a tool, named NERO, that annotates contents and detects smells in NL feature requests. For content annotation, NERO uses a rule matcher developed in their previous work~\cite{ShiCWLB17} to classify a sentence into six categories (e.g., intent and explanation). Using NLP techniques, such as POS tagging, regular expression, and lemmatization, NERO detects 10 smells: vagueness, weakness, generality, coordination ambiguity, referential ambiguity, passive voice, missing condition, missing description, unreadability, and partial content. To evaluate NERO, they applied it to 10 feature requests collected from an issue tracking system.

\citet{SekiHS19} developed a technique for detecting smells in use case descriptions. They defined a catalog of smells based on seven smell characteristics and five smell scopes. The seven smell characteristics are ambiguity, incorrectness, granularity, redundancy, lack, misplacement, and inconsistency. The five smell scopes are use case, section, flow, sentence, and word. \citet{SekiHS19} analyzed 30 use case descriptions written in Japanese to define the smell catalog. They applied Goal-Question-Metric (GQM) paradigm~\cite{Basili94} to automatically detect a subset of their smell catalog and evaluated their prototype tool using eight use case descriptions written in Japanese.

In contrast to these smell detection techniques, \TOOL suggests appropriate requirements patterns (based on Rimay) to fix any detected smells in an NL requirement and thus improve the quality of the requirement. In addition, our work relies on a large set of 2725 information system requirements obtained from a financial company.
Our experiment results show that \TOOL is accurate in detecting smells with a precision and recall of 89\%. It is worth noting that existing techniques~\cite{FemmerFWE17,FerrariGRTBFG18} for detecting requirements smells achieved significantly less accurate experimental results. Because of their different focus, \citet{MuSZZZ20} and \citet{SekiHS19} evaluated their techniques on ten feature requests and eight use case descriptions, respectively.
We defined our smell catalog based on an analysis of Rimay and 1404 requirements, and then evaluated \TOOL on the remaining 1321 requirements. We also note that 6 out of 9 smells detected by \TOOL are not addressed by any of the existing works. These smells are  ``Incomplete System Response'', ``Incomplete Condition'', ``Not Requirement'', ``Incorrect Order Requirement'', ``Incomplete Requirement'', and ``Non-atomic''. These smells violate quality attributes --- completeness, clarity, atomicity, and correctness --- that Rimay enforces in writing requirements (see Section~\ref{sec:smells and patterns}).

Since we defined our smell catalog by analyzing Rimay's concepts and constructs, \TOOL can only detect smells that violate recommendations and best practices guided by Rimay, a language that was defined by qualitatively analyzing information system requirements' needs and practices. Though this approach enables us to provide recommendation as Rimay patterns, we acknowledge that other smells (e.g., non-verifiable terms supported by Smella~\cite{FemmerFWE17}) or other types of ambiguities (e.g., attachment and analytic ambiguities~\cite{OsamaZAGI20}) are not supported. However, one can easily combine these approaches with \TOOL and get the combined benefits of all these approaches.

%% file: conclusions.tex
\section{Conclusions}
\label{sec:qa_conclusions}

The goal of our work is to better support business analysts in the specification of natural language (NL) requirements by detecting smells in them and by guiding the fixing of such smells. To achieve these objectives, we propose a set of nine smells that represent the most common syntactic and semantic errors found in NL requirements from financial applications. Furthermore, we derived 10 patterns aiming at fixing the smells present in NL requirements and converting NL requirements into requirements expressed in Rimay, a controlled natural language (CNL) that was recently proposed to help define unambiguous and complete requirements. We then devised an automated approach to detect our proposed smells and suggest Rimay patterns to improve the quality of requirements.

We evaluated \TOOL in a large industrial case study involving 13 system requirements specifications (SRSs) from information systems in the financial domain, containing 2725 human-annotated NL requirements. This evaluation measured the performance of \TOOL in detecting smells and suggesting accurate Rimay patterns.
Our experiment results show that \TOOL detected smells  with a precision and a recall of 89\%. Furthermore, \TOOL suggested Rimay patterns with a precision of 96\% and a recall of 94\%.
Such patterns help the analyst identify what requirement segments are missing, warrant change, or must be re-ordered.



\revision{
In future work, we intend to expand our list of smells to provide broader coverage of smell detection.
Our proposed smells, associated with the quality attributes enforced by Rimay, identify common problems found in the NL requirements of financial applications, which in all likelihood are not specific to that domain.
However, they may not represent all syntactic and semantic errors present across all NL requirements.
We plan to account for other quality attributes (e.g., comprehensibility and feasibility)~\cite{Denger2005,MontgomeryFBSM22} to identify and rectify relevant requirement smells.}
\revision{Furthermore, it would be important to conduct a user study to assess the economic benefits that organizations and analysts might reap from integrating \TOOL into their requirements engineering process.
This would provide a more holistic view of \TOOL's utility and potential return on investment for its users in the development of their IT systems.}
\revision{Lastly, interesting research directions include the use of chatbot interfaces and large language models (LLMs) in \TOOL.
A chatbot interface would allow analysts to query and receive explanations for detected smells in real-time, thereby facilitating the rectification of identified issues through conversational guidance.
Furthermore, the use of LLMs could improve \TOOL's comprehension of NL requirements, enabling the detection of additional quality issues and the suggestion of corresponding recommendations to address them.}



